\documentclass[12pt]{article}
\usepackage{bm}
\usepackage{graphicx}
\usepackage{verbatim,color, array}
\usepackage{caption}
\usepackage{rotating}
\usepackage{pdflscape}
\usepackage{multirow}
\usepackage{multicol}
\usepackage{amsthm, amssymb}
\usepackage{float}
\usepackage{amsmath}
\usepackage{booktabs}
\usepackage{natbib}

\newcommand{\bq} {\boldsymbol{q}}

\newcommand{\stackiid}{\stackrel{{iid}}{\sim}}

\def\boxit#1{\vbox{\hrule\hbox{\vrule\kern6pt
          \vbox{\kern6pt#1\kern6pt}\kern6pt\vrule}\hrule}}

\def\bse{\begin{eqnarray*}}
\def\ese{\end{eqnarray*}}
\def\be{\begin{eqnarray}}
\def\ee{\end{eqnarray}}
\def\bq{\begin{equation}}
\def\eq{\end{equation}}
\def\bse{\begin{eqnarray*}}
\def\ese{\end{eqnarray*}}

\usepackage{authblk}
\title{Comparison of Small Area Procedures based on Gamma Distributions with Extension to Informative Sampling}

\author[]{Yanghyeon Cho}
\author[]{Emily Berg}
\affil[]{Department of Statistics, Iowa State University, Ames, IA 50011, USA}
\date{}

\begin{document}
\maketitle

{\it Abstract: } The gamma distribution is a useful model for small area prediction of a skewed response variable. We study the use of the gamma distribution for small area prediction. We emphasize a model, called the gamma-gamma model, in which the area random effects have gamma distributions. We compare this model to a generalized linear mixed model. Each of these two models has been proposed independently in the literature, but the two models have not yet been formally compared. We evaluate the properties of two mean square error estimators for the gamma-gamma model, both of which incorporate corrections for the bias of the estimator of the leading term. Finally, we extend the gamma-gamma model to informative sampling. We conduct thorough simulation studies to assess the properties of the alternative predictors. We apply the proposed methods to data from an agricultural survey. 

\newpage

\section{Introduction}

\hspace{.2 in} Asymmetric, positive data occur widely in real-world applications. Examples documented in the fields of health, economics, and agriculture include the body mass index (\cite{pfeffermann2007small}), poverty-related measurements (\cite{molina2010small}), and sheet and rill erosion (\cite{berg2014small}). These types of data are often used to gain a deeper understanding of the characteristics of sub-populations (sub-domains) defined by geographic regions or socio-demographic groups. Such subdivisions are usually more granular than planned estimation domains and therefore have sample sizes that are small or even zero. This motivates the definition of a small area (domain) as any sub-population where the area-specific data are insufficient to assure direct domain estimates of acceptable precision. Estimation procedures for small domains commonly employ indirect estimators based on small area models that incorporate between-area variation and auxiliary variables  \citep{rao2015small, jiang2006mixed, pfeffermann2013new, morales2020small}. A fundamental small area model is the unit-level linear mixed model of \cite{battese1988error}. This model assumes normal distributions and is not immediately suitable for positive, skewed data. 

\hspace{.2 in }  A common approach for skewed data is to apply the unit-level linear mixed model of \cite{battese1988error}, after an appropriate transformation. In the framework of a unit-level lognormal model, \cite{berg2014small} develop closed-form expressions for an empirical Bayes predictor of a small area mean. \cite{lyu2020empirical} and \cite{zimmermann2018small} extend the lognormal model to zero-inflated data and informative sampling, respectively. \cite{berg2014small}, \cite{lyu2020empirical}, and \cite{zimmermann2018small} focus on prediction of means, but many small area parameters are more complex functions of the model response variable. \cite{molina2010small} obtain a Monte Carlo approximation for the best predictor of a general parameter, assuming that transformed study variables follow the nested error linear regression model. \cite{guadarrama2018small} develops predictors of general parameters under an informative sample design.  \cite{rojas2020data} extend \cite{molina2010small} to data-driven transformations that are more general than the log transformation. The use of a transformation (such as a log transformation) is often undesirable. The transformation of the data to a normal distribution can be artificial. In addition, the interpretation of results based on transformed data is difficult.  

\hspace{.2 in} An alternative to a transformation is to model study variable directly. The gamma distribution enables the analyst to model skewed data in the original scale, without need for a transformation.  \cite{hobza2020small} compares several small area predictors, developed under a generalized linear mixed model (GLMM) with a gamma response distribution. A challenge with the gamma GLMM is that the likelihood involves intractable integrals.  \cite{hobza2020small} estimate the model parameters by maximizing the Laplace approximation to the log-likelihood, but this procedure may perform poorly if the gamma distribution differs from a normal distribution.  \cite{dreassi2014small} use Bayesian inference procedures to construct small area estimates under the assumptions of a gamma GLMM. \cite{berg2016small} compare predictors based on lognormal and gamma distributions through simulation, and find that the predictors based on the gamma distribution are more robust to model misspecification. \cite{Graf2019} develop empirical best small area predictors of both means and more general parameters under the assumptions of a generalized gamma inverse-gamma distribution. Unlike the gamma GLMM, the model of \cite{Graf2019} leads to tractable integrals and predictors with closed-form expressions.   The works of \cite{hobza2020small} and \cite{Graf2019} provide the impetus for the research in this paper.  

\hspace{.2 in} We study small area predictors based on gamma distributions. We focus on a unit-level model with a gamma response distribution and gamma distributed random effects, which we call the ``gamma-gamma'' model. The gamma-gamma model is a special case of the more general model of \cite{Graf2019}, with a slightly different parametrization. We compare predictors based on the gamma-gamma model to the predictors based on the gamma GLMM of \cite{hobza2020small}. We develop predictors for the gamma-gamma model in the context of an informative sample design. Our approach to informative sampling transfers the fundamental concepts of \cite{pfeffermann2007small} to the gamma-gamma framework. 
 
\hspace{.2 in } Although the models in this paper are not new, our work has several important contributions. First, the extension of the gamma-gamma model to informative sampling is the most substantive contribution because \cite{Graf2019} only consider noninformative designs.  Second, we formally compare predictors based on the gamma-gamma model to predictors based on the gamma GLMM through simulation. \cite{Graf2019} only compare their model to a lognormal model, and \cite{hobza2020small} exclusively consider the gamma GLMM. Our third contribution is in the area of MSE estimation. We evaluate the properties of several MSE estimators for the gamma-gamma model through simulation. We also propose an MSE estimator that has not yet been used in combination with the gamma-gamma model.  \cite{Graf2019} propose an MSE estimator but do not evaluate its properties through simulation. Our final two contributions are relatively minor but are valuable nonetheless. We develop predictors for the gamma-gamma model using a hierarchical formulation that is computationally easier to implement than the formulation based on marginal distributions in \cite{Graf2019}. Finally,  we generalize the procedures of \cite{hobza2020small} slightly to prediction of parameters that are more general than the class of additive parameters. \cite{hobza2020small} define predictors for additive parameters of the form $N_{i}^{-1}\sum_{j = 1}^{N_{i}} \tilde{h}(y_{ij})$,  where $y_{i1},\ldots, y_{i N_{i}}$ denote the variables of interest for the $N_{i}$ elements of the population for area $i$. We define predictors for more general parameters of the form $h(y_{i1},\ldots, y_{i N_{i}})$. An important type of non-additive parameter that we consider in our study is the population quantile.   

The procedures discussed in this paper are relevant to studies of sheet and rill erosion, or soil loss due to the flow of water. Small area estimates of sheet and rill erosion are valuable for assessing the efficacy of conservation programs.  Sheet and rill erosion is positive, and past studies have documented the distribution of sheet and rill erosion to be skewed right.  We apply the methods detailed in this paper to construct small area estimates of sheet and rill erosion using data from a national survey of cropland in the United States.

\hspace{.2 in} Our study of small area prediction based on gamma distributions is organized as follows.  In Section 2, we define the gamma-gamma model and the gamma GLMM. In Section 3, we propose two MSE estimators for the gamma-gamma model. The first MSE estimator differs from the MSE estimator of \cite{Graf2019}. The second applies a bootstrap bias correction to the MSE estimator of \cite{Graf2019}.   In Section 4, we extend the gamma-gamma model to an informative sample design. In Section 5, we present three simulation studies that (1) compare the gamma-gamma model to the gamma GLMM, (2) evaluate the alternative MSE estimators, and (3) study the properties of the predictors under varying degrees of informativeness.  The data analysis is presented in Section 6. We summarize the main conclusions in Section 7. 

\section{Small Area Estimation based on Gamma Distributions}

\hspace{.2 in} We establish a common notation that we will use for both the gamma-gamma model and the gamma GLMM. Let $i = 1,\ldots, D$ denote the areas, and let $j= 1,\ldots, N_{i}$ index the elements in the population for area $i$. Let $y_{ij}$ denote the response variable for unit $j$ in area $i$, where the support of $y_{ij}$ is $(0,\infty)$. Let $\bm{x}_{ij}$ denote the covariate associated to element $(i,j)$. Assume that $y_{ij}$ is observed for a sample of $n_{i}$ elements in area $i$. Without loss of generality, let $j = 1,\ldots, n_{i}$ index the sampled elements, and let $j = n_{i} + 1,\ldots, N_{i}$ index the non-sampled elements. We let $\bm{y}_{i} = (\bm{y}_{is}',\bm{y}_{ir}')'$, where $\bm{y}_{is} = (y_{i1},\ldots, y_{i n_{i}})'$, and $\bm{y}_{ir} = (y_{i n_{i}+1},\ldots, y_{i N_{i}})'$.  Assume the covariate $\bm{x}_{ij}$ is known for all elements of the population. We consider prediction of a general parameter defined as 
\begin{align}\label{def_parameter}
    \theta_{i} = h(y_{i1}, \ldots, y_{i N_{i}}), 
\end{align}
where $h(\cdot)$ is a specified function.

Suppose a population model is specified for $y_{ij}$. Under the model, the minimum mean square error predictor of $\theta_{i}$ is defined as 
\begin{align*}
\tilde{\theta}_{i}^{BP} = E_{p}(\theta_{i} \mid \bm{y}_{is}),
\end{align*}
which is called the best predictor (BP) of the area parameter.  In this section, we develop   best predictors of $\theta_{i}$ under two models that assume a gamma distribution for the response variables. Section 2.1 and Section 2.2 describe the gamma-gamma model and the gamma, GLMM  respectively. 



\subsection{Unit-Level Gamma-Gamma Small Area Model}

Assume that the  population is generated under a gamma-gamma small area model defined as 
\begin{align}\label{Gam-Gam}
    y_{ij}\mid u_{i} &\stackrel{ind}{\sim}\text{Gamma}(\alpha,\eta_{ij}), \text{\hspace{0.35cm} $j=1,\hdots,N_{i}$,   $i=1,\hdots,D$,}, 
\end{align}
where  $\eta_{ij} = \exp(\bm{x}_{ij}^{T}\bm{\gamma})u_{i},$   $u_{i} \stackrel{iid}{\sim} \text{Gamma}(\delta,\delta),$   and $\bm{\gamma}_{1} = (\gamma_{0},\hdots,\gamma_{p+1})^{T}$.  We use the notation $\text{Gamma}(a,b)$ to denote a gamma distribution with shape parameter $a$ and rate parameter $b$. The model (\ref{Gam-Gam}) can be viewed as a special case of \cite{Graf2019}, with a slightly different parametrization. We use a gamma distribution for $u_{i}$, while \cite{Graf2019} use an inverse-gamma distribution for the area random effect in a transformed scale.   

We first consider small area prediction of the mean defined as
\begin{align*}
    \bar{y}_{N_i} = \frac{1}{N_i}\sum_{j=1}^{N_i}y_{ij}, \text{\hspace{0.1in}$i=1,\hdots,D$.}
\end{align*}
{\it Theorem 1} gives the best predictor of $\bar{y}_{N_{i}}$ under the model (\ref{Gam-Gam}). Although {\it Theorem 1} can be cast as a special case of results in \cite{Graf2019}, we state {\it Theorem 1} and its proof here for completeness. The formulas in {\it Theorem 1} are slightly different than the formulas in \cite{Graf2019} because we parameterize the distribution of the random effect differently. 

{\it Theorem 1:} Under the gamma-gamma model, the best predictor of the small area mean as 
  \begin{align}\label{Pred_mean}
      \tilde{\bar{y}}_{N_i}^{BP}(\bm{\gamma}, \alpha, \delta, \bm{y}_{is})&= \frac{1}{N_i}\Bigg[\sum_{j=1}^{n_i}y_{ij}+\sum_{j=n_{i}+1}^{N_i}\alpha \exp(-\bm{x}_{ij}^{T}\bm{\gamma})\frac{\big\{\sum_{j=1}^{n_{i}}y_{ij}\exp(\bm{x}_{ij}^{T}\bm{\gamma})\big\}+\delta}{n_{i}\alpha+\delta-1}\Bigg]. 
  \end{align}
A proof of {\it Theorem 1} is given in Appendix A. We use the notation ${\bar{y}}_{N_i}^{BP}(\bm{\gamma}, \alpha, \delta, \bm{y}_{is})$ to emphasize dependence of the best predictor on the unknown model parameters and the observed data.

\hspace{.2 in }  We next consider prediction of more general parameters of the form (\ref{def_parameter}). Depending on the complexity of the real-valued function $h(\cdot)$ in   equation (\ref{def_parameter}), an analytic expression of the best predictor, such as the mean predictor in the equation (\ref{Pred_mean}), may not exist. We adopt the approach of \cite{molina2010small} and use a Monte Carlo approximation for the best predictor.   By the proof of Theorem 1, the conditional distribution of $u_{i}$ given the data is given by 
\begin{align*}
    u_{i} \mid \bm{y}_{is} \sim \mbox{Gamma}(\alpha + \delta, \sum_{j =1 }^{n_{i}} y_{ij}\mbox{exp}(\bm{x}_{ij}'\bm{\gamma}) + \tau). 
\end{align*}
This convenient form for the distribution of $u_{i}$ enables us to develop a simple algorithm for approximating the best predictor of $\theta_{i}$. For $\ell = 1,\ldots, L$, repeat the following steps:
\begin{enumerate}
\item Generate $u_{i}^{(\ell)} \sim {\rm Gamma}(n_{i}\alpha+\delta,\sum_{j=1}^{n_{i}}y_{ij}\exp(\bm{x}_{ij}^{T}\bm{\gamma})+\delta)$, $i=1,\hdots,D$.

\item Generate $y_{ij}^{*(r)}\sim {f}(y_{ij}|u_i^{(\ell)};\bm{x}_{ij})$, $j=n_{i}+1,\hdots N_i$.
\item Define 
\begin{align*}
{\theta}_{i}^{(\ell)} = h(\bm{y}_{is}',y_{in_{i+1}}^{*(\ell)}, \ldots, y_{iN_{i}}^{*(\ell)}).
\end{align*}
\end{enumerate}
An approximation for the best predictor of the area parameter is defined as 
\begin{align}\label{BP}
\tilde{\theta}_{i}^{BP}(\bm{\gamma}, \alpha, \delta)  =  \frac{1}{L}\sum_{
    \ell=1}^{L}{\theta}_{i}^{(\ell)}. 
\end{align}
The notation $\tilde{\theta}_{i}^{BP}(\bm{\gamma}, \alpha, \delta)$ emphasizes dependence of the best predictor on the unknown $\bm{\gamma}$, $\alpha$, and $\delta$. 

The algorithm above is slightly simpler than the algorithm of \cite{Graf2019}. We exploit the convenient form of the conditional distribution of $u_{i}$ to generate $y_{ij}$ for nonsampled elements through a hierarchical process that involves first generating $u_{i}$ and then generating $y_{ij}$ given $u_{i}$. In contrast, \cite{Graf2019} generate $y_{ij}$ from the marginal distribution of $y_{ij} \mid \bm{y}_{is}$. Simulating from the conditional distributions, as in our algorithm, is easier than simulating from the marginal distribution, as in \cite{Graf2019}.

The best predictor is a function of the unknown model parameters,  denoted as $\bm{\psi} = (\alpha, \delta,\bm{\gamma}^{T})^{T}$. Calculation of a predictor requires an estimator of the model parameters. We propose to use maximum likelihood estimation. {\it Theorem 2} gives the closed-form expression for the likelihood. 

{\it Theorem 2}:  The likelihood for the model parameters, $\bm{\psi} = (\alpha, \delta,\bm{\gamma}^{T})^{T}$, under the model (\ref{Gam-Gam}) is of the form
\begin{align}\label{likelihood}
L(\bm{\psi}; \bm{y}_{is}) = \prod_{i=1}^{D} f(\bm{y}_{is}\mid\bm{\psi}),
\end{align}
where
\begin{align*}
    f(\bm{y}_{is}\mid \bm{\psi}) &=\frac{\delta^{\delta}}{\{\Gamma(\alpha)\}^{n_{i}}\Gamma(\delta)}\prod_{j=1}^{n_i}y_{ij}^{\alpha-1}\exp\bigg(\alpha (\sum_{j=1}^{n_i}\bm{x}_{ij})^{T}\bm{\gamma}\bigg)\frac{\Gamma(n_{i}\alpha+\delta)}{\bigg(\sum_{j=1}^{n_{i}}\{y_{ij}\exp(\bm{x}_{ij}^{T}\bm{\gamma})\}+\delta\bigg)^{n_{i}\alpha+\delta}}.
\end{align*}
A proof of {\it Theorem 2} is given in Appendix B. Let $\hat{\bm{\psi}} = (\hat{\alpha}, \hat{\delta}, \hat{\bm{\gamma}}')'$ denote the maximum likelihood estimator defined as 
\begin{align*}
\hat{\bm{\psi}} = argmax_{\bm{\psi}} L(\bm{\psi}; \bm{y}_{is}).
\end{align*}
Given the maximum likelihood estimator, we define an empirical Bayes predictor by substitution of $\hat{\bm{\psi}}$ with $\bm{\psi}$. The empirical best predictor of the mean is defined as 
\begin{align}\label{ebhatclsd}
\hat{\bar{y}}_{N_{i}} = \tilde{\bar{y}}_{N_{i}}(\hat{\alpha}, \hat{\beta}, \hat{\bm{\gamma}}, \bm{y}_{is}). 
\end{align}
The predictor (\ref{ebhatclsd}) is obtained by evaluating the closed-form expression for the best predictor of the mean in (\ref{Pred_mean}) at the maximum likelihood estimators. To define an empirical best predictor of a general parameter, we repeat steps 1-3 above with the maximum likelihood estimators in place of the true parameters. This enables us to define the empirical best predictor of a general parameter by 
\begin{align}\label{ebhatgamgam}
\hat{\theta}_{i} = \tilde{\theta}_{i}(\hat{\alpha}, \hat{\beta}, \hat{\bm{\gamma}}, \bm{y}_{is}) .
\end{align}
We refer to the predictor (\ref{ebhatgamgam}) as the EB predictor. When we use the EB predictor (\ref{ebhatgamgam}) to predict the mean, we obtain a MC approximation for the closed-form predictor (\ref{ebhatclsd}).

\subsection{Gamma GLMM}


Define a unit-level gamma GLMM by 
 \begin{align}\label{Hobza}
        y_{ij}\mid v_{i} &\stackrel{ind}{\sim} \text{Gamma}(\nu, \frac{\nu}{\mu_{ij}}), \text{  $i = 1,\hdots, D,$ $j = 1, \hdots, n_{i}$},
    \end{align}
where $g(\mu_{ij}) =\bm{x}_{ij}^{T}\bm{\beta} + v_{i},$ and  $v_{i}\stackrel{iid}{\sim} N(0, \phi^{2})$.  We use the log link function for the mean parameter $\mu_{ij}$, such that $g(\mu_{ij}) = \mbox{log}(\mu_{ij})$.  This model is comparable to the gamma-gamma model in that it has a constant same shape parameter and the area random effects are modeled with only one parameter. Let $\bm{\psi}^{\rm GLMM} = (\bm{\beta}', \phi, \nu)$ denote the parameters of the gamma GLMM. As mentioned in \cite{hobza2020small}, one can fit the model (\ref{Hobza}) using the {\tt R} function {\tt glmer} from the {\tt lme4} package. Let $\hat{\bm{\psi}}^{\rm GLMM} = (\hat{\bm{\beta}}', \hat{\phi}, \hat{\nu})'$ denote the resulting estimates. 

\cite{hobza2020small} proposed three types of predictors for unit-level GLMMs, with emphasis on the gamma GLMM (\ref{Hobza}).  \cite{hobza2020small} restricts attention to additive parameters of the form $N_{i}^{-1}\sum_{j = 1}^{N_{i}} \tilde{h}(y_{ij})$.  We slightly modify their procedures for the purpose of constructing predictors of more general parameters with the form (\ref{def_parameter}) that are not necessarily additive. The first predictor is called an EBP predictor.   Note that, for $i=1,\hdots,D$,
\begin{align}\label{Hobza EBP}
E(\theta_{i}\mid\bm{y}_{is}) 
&= \int h(\bm{y}_{is}, \bm{y}_{ir})f(\bm{y}_{ir}\mid \bm{y}_{is})d\bm{y}_{ir} \nonumber\\
&=\frac{\int h(\bm{y}_{is}, \bm{y}_{ir})\int f(\bm{y}_{ir}\mid {v}_{i})f(\bm{y}_{is}\mid v_{i})f(v_{i})d{v}_{i}d\bm{y}_{ir}}{\int f(\bm{y}_{is}\mid v_{i})f(v_{i})dv_{i}} \nonumber \\
&=\frac{\int \bigg\{\int h(\bm{y}_{is},\bm{y}_{ir})f(\bm{y}_{ir}\mid v_{i})d\bm{y}_{ir}\bigg\}f(\bm{y}_{is}\mid v_{i})f(v_{i})dv_{i}}{\int h(\bm{y}_{is},\bm{y}_{ir})f(\bm{y}_{ir}\mid v_{i})d\bm{y}_{ir}}. 
\end{align}
Using  equation (\ref{Hobza EBP}), we  define the EBP under the model (\ref{Hobza}) using procedures similar to those suggested by Hobza et al. (2020). The iterative Monte Carlo algorithm is as follows:
    \begin{enumerate}
        \item For $\ell_{1}=1,\hdots,L_{1},$ generate $v_{i}^{(\ell_{1})}\sim N(0, \hat{\phi}^{2})$.  
        \begin{enumerate}
            \item for $\ell_{2}=1,\hdots,L_{2},$ generate $\bm{y}_{ir}^{(\ell_{1},\ell_{2})}\sim \hat{f}(\bm{y}_{ir}\mid v_{i}^{(\ell_{1})};\hat{\bm{\psi}}^{\rm GLMM}),$ where
            \begin{align*}
             \hat{f}(\bm{y}_{ir}\mid v_{i}^{(\ell_{1})};\hat{\bm{\psi}}^{\rm GLMM}) =  \prod_{j = n_{i}+1}^{N_{i}} g_{ij}(y_{ij} \mid v_{i}^{(\ell_{1})}, \hat{\bm{\beta}}, \hat{\nu}),
            \end{align*}
            and $ g_{ij}(y_{ij} \mid v_{i}^{(\ell_{1})}, \hat{\bm{\beta}}, \hat{\nu})$ is the density of a gamma distribution with shape parameter $\hat{\nu}$ and rate parameter $\hat{\nu}/\mbox{exp}(\bm{x}_{ij}'\hat{\bm{\beta}} + v_{i}^{(\ell_{1})})$.  
            \item Calculate
             \begin{align*}
                 \hat{A}_{hi}^{(\ell_1)}&=\hat{f}(\bm{y}_{is}\mid v_{i}^{(\ell_{1})};\hat{\bm{\psi}}^{\rm GLMM})\times \frac{1}{L_{2}}\sum_{\ell_{2}=1}^{L_{2}}h(\bm{y}_{is},\bm{y}_{ir}^{(\ell_{1},\ell_{2})})
                 \end{align*}                 
                 and $\hat{B}_{hi}^{(\ell_1)} = \hat{f}(\bm{y}_{is}\mid v_{i}^{(\ell_{1})};\hat{\bm{\psi}}^{\rm GLMM})$, where $\hat{f}(\bm{y}_{is}\mid v_{i}^{(\ell_{1})};\hat{\bm{\psi}}^{\rm GLMM}) = \prod_{j = 1}^{n_{i}} g_{ij}(y_{ij} \mid v_{i}^{(\ell_{1})}, \hat{\bm{\beta}}, \hat{\nu})$.
            \item Approximate the EBP of $\theta_{i}$ as 
            \begin{align}\label{HZ_EBP}
                \hat{\theta}_{i}^{\rm EB\_HZ} = \frac{\sum_{\ell_{1}}^{L_{1}}\hat{A}_{hi}^{(\ell_1)}}{\sum_{\ell_{1}}^{L_{1}}\hat{B}_{hi}^{(\ell_1)}}.
            \end{align}
        \end{enumerate}
\end{enumerate}
The second predictor, called the plug-in predictor, is defined as 
    \begin{align}\label{HZ_PI}
        \hat{\theta}_{i}^{\rm PI} &=  h(\bm{y}_{is}, \tilde{\bm{\mu}}_{ir}),
    \end{align}
    where $\tilde{\bm{\mu}}_{ir} = (\tilde{{\mu}}_{in_{i}+1},\hdots,\tilde{{\mu}}_{iN_{i}})^{T}$, $\tilde{{\mu}}_{ij}= \exp(\bm{x}_{ij}^{T}\hat{\bm{\beta}}+\hat{v}_{i})$, and $\hat{v}_{i}$ is the predicted random effect obtained using \texttt{ranef}.
The last predictor, called the marginal predictor, is defined as 
     \begin{align}\label{HZ_M}
        \hat{\theta}_{i}^{\rm M} &= \frac{1}{L} \sum_{\ell =1}^{L}h(\bm{y}_{is},\bm{y}_{ir}^{M(\ell)}),
    \end{align}
    where $\bm{y}_{ir}^{M(\ell)}\sim f(\bm{y}_{ir}\mid \hat{v}_{i};\hat{\bm{\psi}}^{\rm GLMM})$ for $\ell = 1,\hdots,L.$

\section{MSE Estimation}

In this section, we define two estimators of the MSE of $\hat{\theta}_{i}$. The MSE estimator of Section 3.1 is an adaptation of the general procedure of \cite{https://doi.org/10.48550/arxiv.2210.12221} to the small area context. In Section 3.2, we explain an existing parametric bootstrap MSE estimator in the context of the gamma-gamma model. We later compare the two MSE estimators through simulation in Section 5.2. We do not investigate MSE estimation for the gamma GLMM because we find, through the simulations of Section 5.1, that the gamma-gamma model is generally preferable to the gamma GLMM. 

\subsection{Proposed MSE Estimator}

Suppose we use $L=\infty$ in the prediction procedure of Section 2.1  in order to ignore the variability from the MC approximation used to construct the EB predictor (\ref{ebhatgamgam}). Then, note that the MSE of the predictor $\hat{\theta}_{i}$ can be decomposed as 
\begin{equation}
\begin{aligned}\label{msedecomp1}
{\rm MSE}(\hat{\theta}_{i} ) = M_{i1}+M_{i2},
\end{aligned}
\end{equation}
where $M_{i1} =E[V(\theta_i\mid\bm{y}_{is}; \bm{\psi})] $,  $M_{i2} = E[(\hat{\theta}_{i,\infty}  - \tilde{\theta}_{i,\infty}(\alpha, \delta, \bm{\gamma}) )^{2}]$, and $(\hat{\theta}_{i,\infty}, \tilde{\theta}_{i,\infty}(\alpha, \delta, \bm{\gamma})) = \mbox{lim}_{\ell \rightarrow\infty}(\hat{\theta}_{i}, \tilde{\theta}_{i}(\alpha, \delta, \bm{\gamma}, \bm{y}_{is})$. \cite{https://doi.org/10.48550/arxiv.2210.12221} provide a more rigorous development of the decomposition (\ref{msedecomp1}). Also, see \cite{rao2015small} and \cite{reluga2021simultaneous} for a similar decomposition of the MSE.


The first term $M_{i1}$, called the leading term, is the MSE of the best predictor, and its unbiased estimator is $V(\theta_{i}\mid \bm{y}_{is};\bm{\psi})$. In practice, due to the unknown model parameters $\bm{\psi}$, we use $V(\theta_{i}\mid \bm{y}_{is};\hat{\bm{\psi}})$ as the leading term estimator, and approximate it as 
\begin{equation}
    \begin{aligned}
    \label{leadingtermest}
V(\theta_{i}\mid \bm{y}_{is};\hat{\bm{\psi}}) &\approx  \frac{1}{L-1}\sum_{\ell= 1}^{L}(\hat{\theta}_{i}^{(\ell)} - \hat{\theta}_{i}^{EB})^{2}=:\hat{M}_{1i},   
    \end{aligned}
\end{equation}
where $\hat{\theta}_{i}^{(\ell)}$ are obtained through the EBP procedure in Section 2.1.  

The extra variation induced by replacing ${\bm{\psi}}$ with $\hat{\bm{\psi}}$ in the best predictor is accounted for by the second component $M_{i2}$. The analytical form of $M_{i2}$ is difficult to obtain, so we use the parametric bootstrap to approximate it. For $b = 1,\ldots, B$, repeat the following steps:
\begin{enumerate}
    \item Generate the bootstrap sample $\bm{y}_{is}^{*(b)}= ({y}_{i1}^{*(b)},\hdots, {y}_{in_{i}}^{(b)})^{T} $, $i=1,\hdots,D$ from model (\ref{Gam-Gam}) as
     $  y_{ij}^{*(b)}\stackrel{ind}{\sim}{\rm Gamma}(\hat{\alpha}, \exp(\bm{x}_{ij}^{T}\hat{\bm{\gamma}})u_{i}^{*(b)}) \text{, $j=1,\hdots, n_{i}$, $i=1,\hdots,D$}$,  where $u_{i}^{*(b)}\stackiid {\rm Gamma}(\hat{\delta},\hat{\delta})\text{, $i=1,\hdots,D$}.$ 

    \item Estimate the bootstrap version of the model parameter estimates,  $\hat{\bm{\psi}}^{*(b)}$, by maximizing the likelihood with the bootstrap data generated in step 1. Specifically, $\hat{\bm{\psi}}^{*(b)} = argmax_{\bm{\psi}}L(\bm{\psi}; \bm{y}_{is}^{*(b)})$.  
    
    \item Calculate the bootstrap predictor, $\hat{\theta}_{i}^{*(b)} = \tilde{\theta}_{i}(\hat{\bm{\psi}}^{*(b)}; \bm{y}_{is}^{*(b)})$.  Note that the bootstrap predictor is obtained by applying the algorithm defined in Section 2.1 with the bootstrap parameter estimator and the original data. Implementation of this algorithm results in simulated samples $\theta_{i}^{(\ell, b)}$. Calculate the bootstrap MC approximation for $V\{\theta_{i}\mid\bm{y}_{is};\hat{\bm{\psi}}^{*(b)}\}$, denoted as $\hat{M}_{1i}^{*(b)}$,  as $\hat{M}
_{1i}^{*(b)} = (L-1)^{-1}\sum_{\ell = 1}^{L} (\theta_{i}^{(\ell, b)} - \hat{\theta}
_{i}^{*(b)})^{2}$. 
\end{enumerate}

Then, define the estimator of $M_{2i}$ as:
    \begin{align}\label{M2}
        \hat{M}_{2i} = \frac{1}{B}\sum_{b=1}^{B}(\hat{\theta}_{i}^{EB*(b)} - \hat{\theta}_{i}^{EB})^{2}.
    \end{align}
A preliminary estimator of the MSE of $\theta_{i}$ is be defined as
\begin{align}\label{noBC}
{\rm mse}_{i}^{\rm noBC} = \hat{M}_{1i}+\hat{M}_{2i}.
\end{align}
The label ``noBC'' is used to indicate that the MSE estimator (\ref{noBC}) does not incorporate a correction for the bias of the estimator of the leading term. 

However, the estimator of leading term $\hat{M}_{1i}$ is not an unbiased estimator for $M_{1i}$ due to the replacement ${\bm{\psi}}$ with $\hat{\bm{\psi}}$. To adjust this bias, we may estimate it by utilizing $\hat{M}_{1i}^{*(b)}$, $b=1,\hdots,B$, which is the byproduct of the bootstrap procedure. We can define an additive bias correction as  $\hat{M}_{1i}^{\rm Add} = \hat{M}_{1i}- ({\bar{M}}_{1i}^{*B} - \hat{M}_{1i})$, or a multiplicative correction as, $\hat{M}_{1i}^{\rm Mult} = \hat{M}_{1i}^{2} [\bar{M}_{1i}^{*B}]^{-1}$, where $\bar{M}_{1i}^{*B} = \frac{1}{B}\sum_{b=1}^{B}V\{\theta_{i}\mid\bm{y}_{is};\hat{\bm{\psi}}^{*(b)}\}$. 

These classic additive and multiplicative bias-correction approaches are straightforward, but \cite{hall2006parametric} mention several issues with those approaches. The additive and multiplicative bias-correction could produce a negative leading term estimator when ${\bar{M}}_{1i}^{*B} > \hat{M}_{1i}$ and unreliable estimators, respectively. Thus, \cite{hall2006parametric} suggest a different bias-correction defined as 
\begin{align*}
\hat{M}_{1i}^{\rm HM} &= \begin{cases}
    \hat{M}_{1i}^{\rm Add}, & \hat{M}_{1i}\geq {\bar{M}}_{1i}^{*B},\\
     \hat{M}_{1i}\exp\big[-\{\bar{M}_{1i}^{*B}- \hat{M}_{1i})\big\}/\bar{M}_{1i}^{*B}\big], & \text{if }\hat{M}_{1i}< {\bar{M}}_{1i}^{*B}.
    \end{cases}
\end{align*}
As a special case of the general bias corrections given in \cite{hall2006parametric}, we further define a compromise between $\hat{M}_{1i}^{\rm Add}$ and $\hat{M}_{1i}^{\rm Mult}$ as
\begin{align*}
    \hat{M}_{1i}^{\rm Comp} &= \begin{cases}
    \hat{M}_{1i}^{\rm Add}, & \hat{M}_{1i}\geq {\bar{M}}_{1i}^{*B},\\
      \hat{M}_{1i}^{\rm Mult}, & \text{if }\hat{M}_{1i}< {\bar{M}}_{1i}^{*B}.
    \end{cases}
\end{align*}
In summary, the bias-corrected MSE estimators are constructed by 
\begin{align}\label{HM}
    {\rm mse}_{i}^{\rm HM} = \hat{M}_{1i}^{\rm HM}+\hat{M}_{2i},
\end{align}
and
\begin{align}\label{Comp}
    {\rm mse}_{i}^{\rm Comp} = \hat{M}_{1i}^{\rm Comp}+\hat{M}_{2i}.
\end{align}

\subsection{Existing Parametric Bootstrap MSE Estimators}

Instead of estimating $M_{i1}$ and $M_{i2}$ separately, one can use the parametric bootstrap to estimate ${\rm MSE}(\hat{\theta}_{i})$ directly. \cite{molina2007small}, \cite{Graf2019}, and \cite{hobza2020small} use the single-stage bootstrap method to estimate the MSE. One may implement the double-bootstrap algorithm suggested by \cite{hall2006parametric} to correct the single-stage estimator. However, the double-bootstrap is computationally expensive and may not be feasible for a large population. Thus, we consider a simpler double-bootstrap motivated by \cite{erciulescu2014parametric} and \cite{reluga2021simultaneous}, where we generate only one bootstrap replicate in the second-stage bootstrap.  The following algorithm describes how to obtain those estimators. 
\begin{enumerate}
     \item Obtain the estimate of the model parameter $\hat{\bm{\psi}}$.
     \item For $b_{1} = 1, \hdots, B_{1}$, independently generate the bootstrap population $\bm{y}_{i}^{*(b_{1})}= ({y}_{i1}^{*(b_{1})},\hdots, {y}_{iN_{i}}^{(b_{1})})^{T} $, $i=1,\hdots,D$ from model (\ref{Gam-Gam}) as $ y_{ij}^{*(b_{1})}\stackrel{ind}{\sim}{\rm Gamma}(\hat{\alpha}, \exp(\bm{x}_{ij}^{T}\hat{\bm{\gamma}})u_{i}^{*(b_{1})})$, $j=1,\hdots, N_{i}$, $i=1,\hdots,D$ where  $u_{i}^{*(b_{1})}\stackiid {\rm Gamma}(\hat{\delta},\hat{\delta})$ for $ i=1,\hdots,D$. 
    \item Calculate the bootstrap version of a small area parameter $\theta_{i}^{*(b_{1})}$ with the bootstrap population $\bm{y}_{i}^{*(b_{1})}$, the EBP $\hat{\theta}_{i}^{*(b_{1})}$ with the bootstrap sample $\bm{y}_{is}^{*(b_{1})}$, and $D^{*(b_{1})} = \{\hat{\theta}_{i}^{*(b_{1})}- \theta_{i}^{*(b_{1})}\}^2$.
    \item Estimate the bootstrap model parameter estimates $\hat{\bm{\psi}}^{*(b_{1})} = (\hat{\alpha}^{*(b_{1})},\hat{\delta}^{*(b_{1})},\hat{\bm{\gamma}}^{(b_{1})T})^{T}$ with the bootstrap sample $\bm{y}_{is}^{*(b_{1})}$.
    \begin{enumerate}
        \item For $b_{2} = 1, \hdots, B_{2}$, independently generate the bootstrap population $\bm{y}_{i}^{**(b_{2})}= ({y}_{i1}^{**(b_{2})},\hdots, {y}_{iN_{i}}^{**(b_{2})})^{T} $, $i=1,\hdots,D$ from model (\ref{Gam-Gam}):
    \begin{align*}
        y_{ij}^{**(b_{2})}&\stackrel{ind}{\sim}{\rm Gamma}(\hat{\alpha}^{*(b_{2})}, \exp(\bm{x}_{ij}^{T}\hat{\bm{\gamma}}^{*(b_{1})})u_{i}^{**(b_{1})}) \text{, $j=1,\hdots, N_{i}$, $i=1,\hdots,D$}\\
        u_{i}^{**(b_{2})}&\stackiid {\rm Gamma}(\hat{\delta}^{*(b_{1})},\hat{\delta}^{*(b_{1})})\text{, $i=1,\hdots,D$}.
    \end{align*}
    (We set $B_{2}=1$ to employ the simpler double-bootstrap.)
    \item Calculate the bootstrap version of a small area parameter $\theta_{i}^{**(b_{2})}$ with the bootstrap population $\bm{y}_{i}^{**(b_{2})}$, the EBP $\hat{\theta}_{i}^{**(b_{2})}$ with the bootstrap sample $\bm{y}_{is}^{**(b_{2})}$, and $D^{**(b_{2})} =\{ \hat{\theta}_{i}^{**(b_{2})}- \theta_{i}^{**(b_{2})}\}^2$.
    \item Set ${\rm mse}_{i}^{**(b_1)} = \frac{1}{B_{2}}\sum_{b_{2}=1}^{B_{2}}D^{**(b_{2})}.$ 
    \end{enumerate}
    \item Finally, define the single-stage and the double-bootstrap MSE estimators as 
    \begin{align}
        {\rm mse}_{i}^{S} = \frac{1}{B_{1}}\sum_{b_{1}=1}^{B_{1}} D^{*(b_{1})},
    \end{align}
    and
    \begin{align}
        {\rm mse}_{i}^{D} = 2{\rm mse}_{i}^{S}- \frac{1}{B_{1}}\sum_{b_{1}=1}^{B_{1}}{\rm mse}_{i}^{**(b_1)}.
    \end{align}
\end{enumerate}
Note that the single-stage MSE estimator is comparable to (\ref{noBC}) and the (simpler) double-stage MSE estimator to the proposed bias-corrected MSE estimators.

\section{Extension of SAE Gamma-Gamma model Under Informative Sampling}

We extend the gamma-gamma model to an informative sampling design. We utilize well-known relationships among the population, sample, and sample-complement distributions of $y_{ij}$ established in \cite{pfeffermann2007small}. We assume the same model for the first moment of the sampling weight in \cite{pfeffermann2007small}. 

For completeness, we restate key relationships defined in \cite{pfeffermann2007small} with respect to the second-stage unit $y_{ij}$ and the corresponding sampling weight $w_{ij}$. These are given by
\begin{align}\label{relationship1}
f(y_{ij}\mid \bm{x}_{ij},u_{i},I_{i}=1,I_{ij}=1)&=\frac{E(w_{ij}\mid \bm{x}_{ij}, u_{i},I_{i}=1, I_{ij}=1)}{E(w_{ij}\mid \bm{x}_{ij}, u_{i},y_{ij},I_{i}=1, I_{ij}=1)}f(y_{ij}\mid \bm{x}_{ij},u_{i},I_{i}=1) 
\end{align}
and
\begin{align}\label{relationship2}
    f(y_{ij}\mid \bm{x}_{ij},u_{i},I_{i}=1,I_{ij}=0)&=\frac{E[(w_{ij}-1)\mid \bm{x}_{ij}, u_{i},y_{ij},I_{i}=1, I_{ij}=1]}{E[(w_{ij}-1)\mid \bm{x}_{ij}, u_{i},I_{i}=1, I_{ij}=1]}f(y_{ij}\mid \bm{x}_{ij},u_{i},I_{i}=1,I_{ij}=1),
\end{align}
where $I_{i}$ and $I_{ij}$ are the sample indicators for an area $i$ and unit $j$ in the area $i$, respectively, $w_{ij}=1/P(j \in s_{i})$, and $s_{i}=\{j: I_{ij}=1\}$. These relationships imply that we can deduce adequate information about other distributions from observed units and their weights. For simplicity, we suppose all areas are selected such that $I_{i} = 1$ for $i= 1,\ldots, D$.

For the complex design, we suppose the sample distribution is given by
\begin{align}\label{sampledist}
  y_{ij} \mid \bm{x}_{ij},u_{i},I_{i}=1, I_{ij}=1 &\stackrel{ind}{\sim}  \text{Gamma}(\alpha_{s},\eta_{s,ij})\text{\hspace{0.2cm} $j=1,\hdots,n_{i}$, $i=1,\hdots,D$}, 
  \end{align}
where $ \eta_{s,ij}  = \exp(\bm{x}_{ij}^{T}\gamma_{s})u_{i}\nonumber$ and $u_{i}\stackrel{iid}{\sim} \text{Gamma}(\delta_{s}, \delta_{s})$. Here, the model parameters in (\ref{sampledist}) are differentiated from those of the population distribution (\ref{Gam-Gam}) using the subscript $s$. We further assume that the expected values of the sampling weight satisfies 
\begin{align}\label{sp_wgt}
E(w_{ij}\mid \bm{x}_{ij}, u_{i},y_{ij},I_{i}=1, I_{ij}=1)&=E(w_{ij}\mid \bm{x}_{ij},y_{ij},I_{i}=1,I_{i}=1)\\
&=\kappa_{i}\exp(\bm{x}_{ij}^{T}\bm{a}-by_{ij}),\nonumber
\end{align}
for $b>0$ where $\kappa_{i} = N_{i}^{-1}\sum_{j=1}^{N_{i}}\exp(-\bm{x}_{ij}^{T}\bm{a}+by_{ij})$. Denote the collection of fixed model parameters by  ${\bm{\psi}}^{\rm Info} = (\alpha_{s},\delta_{s},\bm{\gamma}_{s}^{T},\bm{a}^{T},b,\bm{\kappa}^{T})^{T}$, where $\bm{\kappa} = (\kappa_{1}, \hdots, \kappa_{D})^{T}$. Then, using the relationship (\ref{relationship2}), the following sample-complement distributions under the informative sample scheme can be derived as 
\begin{align}\label{fciyij}
f(y_{ij}\mid \bm{x}_{ij}, u_{i}, I_{i}=1, I_{ij}=0; \bm{\psi}^{Info}) 
&= \frac{[E(w_{ij}\mid \bm{x}_{ij},y_{ij},I_{i}=1,I_{ij}=1)-1] f(y_{ij} \mid \bm{x}_{ij},u_{i},I_{i}=1, I_{ij}=1)}{E(w_{ij}\mid \bm{x}_{ij},I_{i}=1,I_{ij}=1)-1} \nonumber \\
&=\frac{\lambda_{ij}}{\lambda_{ij}-1}f(y_{ij}\mid\bm{x}_{ij},u_{i},I_{i}=1)-\frac{1}{\lambda_{ij}-1}f(y_{ij}\mid\bm{x}_{ij},u_{i},I_{i}=1,I_{ij}=1),
\end{align}
where 
\begin{align*}
\lambda_{ij} &= E(w_{ij}\mid\bm{x}_{ij}, I_{i=1}, I_{ij}=1)\\
    &= \kappa_{i}\exp(\bm{a}^{T}\bm{x}_{ij})E[\exp(-by_{ij})\mid\bm{x}_{ij}, u_{i},I_{i=1}, I_{ij}=1]\\
    &= \kappa_{i}\exp(\bm{a}^{T}\bm{x}_{ij})\bigg(1+\frac{b}{\eta_{s,ij}}\bigg)^{-\alpha},
\end{align*}
and the population distribution $f(y_{ij}\mid \bm{x}_{ij},u_{i},I_{i}=1)$ is given by
\begin{align}\label{fp}
f(y_{ij}\mid \bm{x}_{ij},u_{i},I_{i}=1)&\propto E(w_{ij}\mid \bm{x}_{ij}, u_{i},y_{ij},I_{i}=1, I_{ij}=1)f(y_{ij}\mid \bm{x}_{ij},u_{i},I_{i}=1, I_{ij}=1) \text{\hspace{0.2cm}  (\ref{relationship1})}\nonumber\\
&= \frac{(\eta_{s,ij}+b)^{\alpha_{s}}}{\Gamma(\alpha_{s})} y_{ij}^{\alpha_{s}-1}\exp(-y_{ij}(\eta_{ij}+b))
\end{align}
When the observed values are not related to sampling probability (that is, $b=0$), the population and sample-complement distribution, $f(y_{ij}\mid \bm{x}_{ij},u_{i},I_{i}=1)$ and $f(y_{ij}\mid \bm{x}_{ij},u_{i},I_{i}=1, I_{ij}=0)$, are the same as the sample distribution $f(y_{ij}\mid \bm{x}_{ij},u_{i},I_{i}=1, I_{ij}=1)$. In this case, the algorithm for the (empirical) best predictor is identical to that in Section 2.1. 

However, under the informative design, that is $b\neq0$, we need to reflect the informative sampling scheme by using the sample-complement distribution (\ref{fciyij}). The procedure requires an estimator of $\bm{\psi}^{Info}$.  We define $(\hat{\alpha}_{s}, \hat{\delta}_{s}, \hat{\bm{\gamma}}_{s}')'$ to be the maximum likelihood estimator under the sample model.  The estimator of $(\bm{a}^{T},b,\bm{\kappa}_{i})^{T}$ is obtained by minimizing 
\begin{align*}
    {\rm SSE}(\bm{a}^{T},b,\bm{\kappa}^{T}) = \sum_{i=1}^{D}\sum_{j=1}^{n_{i}}\big\{w_{ij} - \kappa_{i}\exp(\bm{x}_{ij}^{T}\bm{a}-by_{ij})\big\},
\end{align*}
as in \cite{pfeffermann2007small}. The procedure for the empirical best predictor under the informative design is then implemented as follows. For $\ell = 1, \ldots, L$, repeat the following steps:
\begin{enumerate}
    \item Generate $u_{i}^{(\ell)} \sim \text{Gamma}(n_{i}\hat{\alpha}_{s}+\hat{\delta}_{s},\sum_{j=1}^{n_{i}}y_{ij}\exp(\bm{x}_{ij}^{T}\hat{\bm{\gamma}}_{s})+\hat{\delta}_{s})$, $i=1,\hdots,D$.

    \item Generate $y_{ij}^{*(\ell)}\sim \hat{f}(y_{ij}\mid \bm{x}_{ij}, u_i^{(\ell)},I_{i}=1,I_{ij}=0; \hat{\bm{\psi}}^{Info})$, $j=n_{i}+1,\hdots N_i$.
    \item Define 
    \begin{align*}
    \hat{\theta}_{i}^{(\ell)} = h(\bm{y}_{is},y_{in_{i+1}}^{*(\ell)}, \ldots, y_{iN_{i}}^{*(\ell)}).
    \end{align*}
\end{enumerate}
Then, the empirical best predictor of the area parameter is defined as 
\begin{align}\label{EBinfor}
\hat{\theta}_{i}^{\rm EB\_INFO}  &= \frac{1}{L}\sum_{
    \ell=1}^{L}\hat{\theta}_{i}^{(\ell)}.
\end{align}
 

{\it Remark 1:} We use inversion sampling to generate $y_{ij}^{*(r)}$ from the sample-complement distribution $f(y_{ij}\mid\bm{x}_{ij},u_{i},I_{i}=1,I_{ij}=0; \hat{\bm{\psi}}^{Info})$ in Step 3 of the procedure. Specifically, we decompose Step 3 into two steps as follows:
\begin{enumerate}
    \item[3.1] Generate $q_{ij}\stackrel{iid}{\sim} \text{Unif}(0,1)$, $j = n_{i+1}.\hdots,N_{i}$.
    \item [3.2] Set $y_{ij}^{*(\ell)} = \hat{F}^{-1}(q_{ij}\mid \bm{x}_{ij}, u_{i}^{(\ell)}, I_{i}=1, I_{ij}=0; \hat{\bm{\psi}}^{Info})$, $j = n_{i+1}.\hdots,N_{i}$, where 
\begin{align*}
    F(y_{ij}\mid \bm{x}_{ij}, u_{i}, I_{i}=1, I_{ij}=0;  \hat{\bm{\psi}}^{Info}) &= \int_{0}^{y_{ij}}f(y_{ij}\mid\bm{x}_{ij}, u_{i}, I_{i}=1, I_{ij}=0)dy_{ij}\\
    &=\frac{\lambda_{ij}}{\lambda_{ij}-1}\frac{\gamma\{\alpha_{s},(\eta_{s,ij}+b)y_{ij}\big\}}{\Gamma(\alpha_{s})}-\frac{1}{\lambda_{ij}-1}\frac{\gamma\{\alpha_{s},\eta_{s,ij}y_{ij}\big\}}{\Gamma(\alpha_{s})},
\end{align*}
and $\gamma(a,x) = \int_{0}^{x}t^{s-1}\mbox{exp}(-t)dt$ is the incomplete gamma function.
\end{enumerate}

{\it Remark 2:} For sufficiently large $E(w_{ij}\mid\bm{x}_{ij}, I_{i=1}, I_{ij}=1)$, one may use the population distribution, $f(y_{ij}\mid \bm{x}_{ij}, u_{i}, I_{i}=1)$, as an approximation for the sample-complement distribution, $f(y_{ij}\mid \bm{x}_{ij}, u_{i}, I_{i}=1, I_{ij}=0)$. Simulating from the population distribution is easier than simulating from the sample complement distribution because the population distribution is a gamma distribution. We use the exact complement distribution because we found that the difference between the population distribution and the complement distribution can be important when the degree of informativeness is large. 



\section{Simulation Study}

We carry out three simulation experiments to evaluate the procedures defined in Sections 2-4. In all simulation studies, we consider $D=100$ areas, each with population size $N_{i} = 100$. We stratify the areas into two strata, where Stratum $H_{1}$ and Stratum $H_{2}$ are composed of areas $1\le i \le 50$ and $51\le i \le D$, respectively. Assign the area sample size of  $n_{i}=10$ if $i\in H_{1}$  and $n_{i}=20$ if $i\in H_{2}$. Samples of size $n_{i}$ are selected independently across the areas using simple random sampling without replacement.  Then, for the population units, we simulate the study variables $y_{ij}$ in conjunction with the values of the auxiliary variable $x_{ij} \sim U(0,2)$ for $j=1,\hdots,N_{i}$, where $x_{ij}$ are held constant throughout MC simulations. The true model for simulating a population is defined in each experiment.

\hspace{.2 in} In addition to the small area mean, we take into account three non-additive parameters. The first two are the  $25$th and $75$th sample quantiles, denoted as $Q_{0.25}$ and $Q_{0.75}$, respectively.   
These are calculated through the function \texttt{quantile} in \texttt{R} with the default method.
The second is the Gini Coefficient (abbreviated Gini) defined as 
	$${\rm Gini}_i = \frac{\sum_{k=1}^{N_i}\sum_{\ell=1}^{N_i}\mid y_{ik} - y_{i\ell}\mid}{2N_i^{2}\bar{Y}_{i}},$$
	where the value is obtained by the function \texttt{gini} of \texttt{R} package \texttt{reldist}.

\subsection{Simulation 1}

The objectives of Simulation 1 are to evaluate the performance of the gamma-gamma predictors in comparison to the three predictors suggested by \cite{hobza2020small} and to examine the robustness of the gamma-gamma predictors against model misspecification. To attain these goals, the values of target variable $y$ are generated from either the gamma-gamma model (\ref{Gam-Gam}) or the GLMM (\ref{Hobza}). 

\subsubsection{The Gamma-Gamma model}

We first generate data from the model (\ref{Gam-Gam}). We set $\bm{\gamma}=(1,0.5)^{T}$, $\delta = 4$, and consider the three values of $\alpha \in \{1,2.5,5\}$. Note that the skewness of the distribution descreases as $\alpha$ increases. 

In each simulation iteration, we compute the considered predictors, $\hat{\theta}_{i}^{(m),\text{ } {\rm pred}}$, ${\rm pred} \in\{\rm EB, EB\_HZ, EB\_M, EB\_PI\}$, where ${\rm EB}$, ${\rm EB\_HZ}$, ${\rm EB\_M}$ and ${\rm EB\_PI}$ are defined in (\ref{BP}), (\ref{HZ_EBP}), (\ref{HZ_PI}), and (\ref{HZ_M}), respectively. For $\bar{Y}$, the closed-form expression   (\ref{Pred_mean}) is also used, and is denoted as EB\_clsd. Further, calculate the direct estimator of each considered parameter, denoted as Dir, which can be obtained by passing the area-specific sample obtained in step 3 as the argument of the area parameter function.


\hspace{0.2 in} We compare the predictors using the relative bias (RB) and relative root MSE (RRMSE). The RB and RRMSE for predictor $pred$ are defined as 
\begin{align*}
{\rm RB}_{i} = \frac{M^{-1}\sum_{m=1}^{M}(\hat{\theta}_{i}^{(m), \rm pred}-\theta_{i}^{(m)})}{M^{-1}\sum_{m=1}^{M}\theta_{i}^{(m)}}, \hspace{0.2cm} \mbox{ and } {\rm RRMSE}_{i} = \frac{\sqrt{M^{-1}\sum_{m=1}^{M}(\hat{\theta}_{i}^{(m), \rm pred}-\theta_{i}^{(m)})^{2}}}{M^{-1}\sum_{m=1}^{M}\theta_{i}^{(m)}}, 
\end{align*}
where $\theta_{i}^{(m)}$ denotes the area population parameter obtained in MC simulation $m$. The averages of RB and RRMSE (in \%) for areas within the same stratum (same sample size) are shown in Table \ref{sim1.1_EB_RB_RE}. 

The $\rm EB$ and $\rm  EB\_clsd$ predictors are superior to the alternatives for this configuration. EB has RB closest to zero for most cases, and the RB of EB is uniformly below 1\% in absolute value. The RRMSE of $\rm EB$ is consistently smaller than the RRMSE of the $\rm EB\_HZ$, $M$, PI, or Dir predictors. This is expected because $\rm EB$ is an estimator of the optimal (minimum MSE) predictor for this simulation model. For the mean, $\rm EB\_clsd$ is more efficient than $\rm EB$, which is expected because $\rm EB$ is an MC approximation for $\rm EB\_clsd$. The loss of efficiency from use of the MC approximation through $\rm EB$, relative to $\rm EB\_clsd$, is slight. 

The properties of the $\rm EB\_HZ$, $M$, and $\rm PI$ predictors reflect the patterns described in \cite{hobza2020small}. The RB of the PI predictor makes an important contribution to the RRMSE for all parameters, except for the mean. This occurs because the PI predictor replaces a non-sampled unit with its estimated conditional mean. Therefore, the $25$th quantile is predicted to be greater than the actual value, while the $75$th quantile is predicted to be lower. A similar phenomenon occurs with the PI predictor of Gini. In terms of RRMSE, $M$ compares favorably to $\rm EB\_HZ$, which is consistent with the simulation results of \cite{hobza2020small}. 

As expected, the efficiency of the direct estimator depends heavily on the sample size. The direct estimator is inefficient for $Q_{0.25}$ and for $\rm Gini$, as a result of the small sample size and the nonlinearity of these parameters. For the mean, the direct estimator is more efficient than $\rm EB\_HZ$. 



\begin{table}[H]
	\centering
	\scalebox{0.75}{
		\begin{tabular}{ccc|ccccc|ccccc}
			\toprule
			\toprule
			\multirow{2}{*}{\large{\bfseries Parameter }} & \multirow{2}{*}{\large{\bfseries $\alpha$}} & \multirow{2}{*}{\large{\bfseries $n_{i}\,$}} &\multicolumn{5}{c}{\large{\bfseries RB (\%) }}&\multicolumn{5}{c}{\large{\bfseries RRMSE (\%) }}\\
			\cmidrule(r){4-13}
			& & &{\large{\bfseries EB}} & {\large{\bfseries $EB\_HZ$}} & {\large{\bfseries $M$}} & {\large{\bfseries PI}} & {\large{\bfseries Dir}} & {\large{\bfseries EB}} & {\large{\bfseries $EB\_HZ$}} & {\large{\bfseries $M$}} & {\large{\bfseries PI}} & {\large{\bfseries Dir}} \\ 
			\midrule
			h1 & 1.00 & 10 & 0.05 (0.05) & -2.02 & -6.59 & -6.59 & 0.02 & 33.48 (33.40) & 38.36 & 34.99 & 34.98 & 40.21 \\ 
			&  & 20 & -0.02 (-0.03) & -1.50 & -3.67 & -3.67 & -0.05 & 23.49 (23.45)& 28.74 & 24.29 & 24.28 & 26.39 \\ 
			& 2.50 & 10 & 0.05 (0.05) & -1.98 & -6.57 & -6.57 & -0.01 & 33.14 (33.07) & 38.26 & 34.93 & 34.92 & 39.60 \\ 
			&  & 20 & 0.04 (0.04) & -1.47 & -3.61 & -3.61 & 0.04 & 23.59 (23.52)& 29.70 & 24.40 & 24.39 & 26.54 \\ 
			& 5.00 & 10 & 0.04 (0.04)& -4.12 & -5.08 & -5.08 & 0.03 & 22.45 (22.39)& 36.39 & 26.30 & 26.30 & 26.39 \\ 
			&  & 20 & 0.02 (0.02) & -3.23 & -2.74 & -2.74 & -0.00 & 15.32 (15.29) & 28.91 & 16.91 & 16.91 & 17.65 \\ 
			\midrule
			h2 & 1.00 & 10 & 0.09 (0.09) & 2.17 & -2.04 & 142.96 & 26.74 & 39.03 & 43.92 & 39.64 & 168.48 & 88.62 \\ 
			&  & 20 & 0.02 & 2.20 & 0.59 & 144.00 & 11.65 & 30.24 & 35.53 & 30.48 & 171.84 & 54.62 \\ 
			& 2.50 & 10 & 0.10 & 2.19 & -2.04 & 142.92 & 26.72 & 38.68 & 43.65 & 39.45 & 167.81 & 88.29 \\ 
			&  & 20 & 0.10 & 2.23 & 0.65 & 144.12 & 11.76 & 30.45 & 36.62 & 30.63 & 172.54 & 55.09 \\ 
			& 5.00 & 10 & 0.04 & -5.34 & -5.89 & 39.17 & 11.16 & 24.88 & 40.10 & 29.07 & 48.72 & 44.31 \\ 
			&  & 20 & 0.04 & -4.45 & -3.40 & 40.89 & 4.96 & 17.98 & 34.12 & 19.76 & 49.48 & 28.40 \\ 
			\midrule
			h3 & 1.00 & 10 & 0.04 & -2.05 & -6.40 & -15.06 & -1.91 & 34.72 & 39.97 & 36.16 & 40.59 & 47.07 \\ 
			&  & 20 & -0.01 & -1.53 & -3.47 & -12.51 & -0.95 & 25.04 & 30.97 & 25.81 & 30.84 & 32.11 \\ 
			& 2.50 & 10 & 0.04 & -2.00 & -6.38 & -15.02 & -1.89 & 34.37 & 39.75 & 36.06 & 40.51 & 46.16 \\ 
			&  & 20 & 0.02 & -1.52 & -3.44 & -12.46 & -0.87 & 25.21 & 32.22 & 26.04 & 31.08 & 32.28 \\ 
			& 5.00 & 10 & 0.03 & -4.08 & -4.95 & -11.26 & -2.34 & 23.40 & 38.14 & 27.16 & 31.26 & 31.41 \\ 
			&  & 20 & 0.01 & -3.22 & -2.60 & -9.05 & -1.09 & 16.51 & 30.94 & 17.94 & 22.26 & 21.84 \\ 
			\midrule
			h4 & 1.00 & 10 & -0.02 & -1.57 & -1.87 & -59.85 & -9.60 & 5.77 & 6.05 & 6.14 & 60.19 & 19.30 \\ 
			&  & 20 & -0.02 & -1.39 & -1.70 & -51.87 & -4.32 & 5.44 & 5.68 & 5.76 & 52.22 & 12.36 \\ 
			& 2.50 & 10 & -0.01 & -1.54 & -1.84 & -59.81 & -9.60 & 5.77 & 6.04 & 6.13 & 60.15 & 19.29 \\ 
			&  & 20 & -0.02 & -1.38 & -1.70 & -51.84 & -4.29 & 5.45 & 5.68 & 5.77 & 52.19 & 12.34 \\ 
			& 5.00 & 10 & -0.02 & 1.04 & 0.68 & -49.16 & -9.58 & 6.40 & 6.57 & 6.47 & 49.62 & 21.13 \\ 
			&  & 20 & -0.01 & 0.97 & 0.57 & -43.25 & -4.29 & 6.04 & 6.26 & 6.09 & 43.71 & 13.65 \\ 
			\bottomrule  \bottomrule
		\end{tabular}
	}
	\caption{RB ($\%$) and RRMSE ($\%$) for EB, EB\_{HZ}, M, PI, and Dir of considered parameters by sample size under the gamma-gamma model. In parentheses are the RB (\%) and RRMSE (\%) values for EB\_clsd.} 
	\label{sim1.1_EB_RB_RE}
\end{table}



\subsubsection{GLMM}

In this subsection, we generate data from the model (\ref{Hobza}) with $\nu = \{1,2.5,5\}$, $\bm{\beta} = (0.5,0.05)^{T}$, and $\phi^{2} =0.1^{2}$. To gain a deeper understanding of the predictors proposed by Hobza et al. (2020), the best predictors, denoted as BP\_HZ and BP\_M, were additionally calculated using the value of the true model parameter. 

\begin{figure}[H]
	\captionsetup{font=footnotesize}
	\centering
	\includegraphics[scale=0.7]{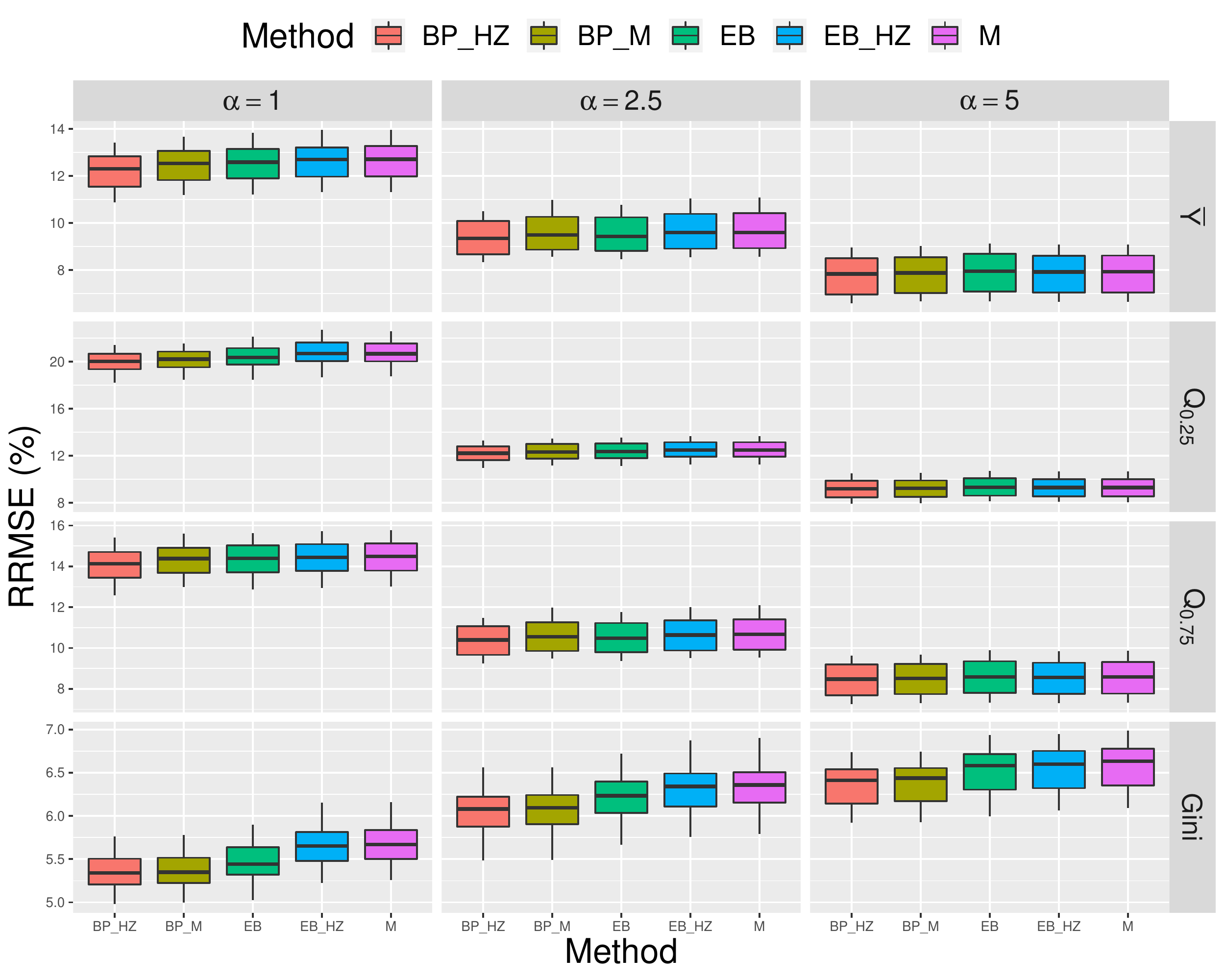}
	\caption{Box plots of RRMSE(\%) for considered predictors under the GLMM. $BP$ indicates the proposed best predictor with true model parameters.}
	\label{RE_GMM}
\end{figure}

What is most interesting about this configuration is that the empirical best predictors for the gamma-gamma model remain competitive, even though the data are generated from the gamma GLMM. For $\nu \in \{1.00, 2.50\}$, the RRMSEs of the EB and EB\_clsd predictors are below the RRMSEs of the alternative predictors. For Gini, the EB predictor has uniformly smallest RRMSE. For the other parameters and $\nu = 5$, the increase in RRMSE from EB, relative to EB\_HZ is slight. As illustrated in Figure~\ref{RE_GMM}, the problems with the EB\_HZ and $M$ predictors occur due to the effect of the variance of parameter estimators. The HZ and $M$ predictors calculated with the true parameters (abbreviated BP\_HZ and BP\_M in Figure~\ref{RE_GMM}) are most efficient in terms of RRMSE for the GLMM configurations. This result has two implications. The first is that estimating the model parameters with the Laplace approximation for the likelihood may not work well when the skewness of the distribution is large. The second is that the empirical best predictors for the gamma-gamma model appear robust to this form of model misspecification. Specifically, the empirical best predictors for the gamma-gamma model maintain reasonable efficiency, even when the data are generated from the gamma GLMM. 

The results for the PI and Dir predictors for the GLMM are similar to the results for these predictors for the gamma-gamma model. The use of the estimated  mean as the predictor for a nonsampled element for the PI predictor causes the PI predictor to have an important bias for nonlinear parameters. The direct estimator is also inefficient for nonlinear parameters as a result of the small area sample size. 


\begin{table}[H]
	\centering
	\scalebox{0.7}{
		\begin{tabular}{ccc|ccccc|ccccc}
			\toprule
			{\large{\bfseries Parameter }} & {\large{\bfseries $\nu$}} & {\large{\bfseries $n_{i}$}} & {\large{\bfseries EB}} & {\large{\bfseries EB\_HZ}} & {\large{\bfseries $M$}} & {\large{\bfseries PI}} & {\large{\bfseries Dir}} & {\large{\bfseries EB}} & {\large{\bfseries EB\_HZ}} & {\large{\bfseries $M$}} & {\large{\bfseries PI}} & {\large{\bfseries Dir}} \\ 
			\midrule
			h1 & 1.00 & 10 & 0.13 (0.13) & -0.26 & -0.83 & -0.83 & -0.07 & 13.20 (13.17) & 13.28 & 13.31 & 13.29 & 30.23 \\ 
			&   & 20 & 0.11 (0.12) & -0.19 & -0.67 & -0.67 & 0.18 & 11.88 (11.86)& 11.98 & 11.99 & 11.98 & 20.31 \\ 
			& 2.50 & 10 & 0.10 (0.10)& -0.38 & -1.02 & -1.02 & 0.12 & 10.24 (10.22) & 10.38 & 10.41 & 10.40 & 19.16 \\ 
			&   & 20 & 0.02 (0.02)& -0.30 & -0.79 & -0.79 & -0.03 & 8.81 (8.79)& 8.93 & 8.94 & 8.93 & 12.75 \\ 
			& 5.00 & 10 & 0.15 (0.14)& -0.34 & -0.79 & -0.79 & 0.03 & 8.71 (8.68)& 8.64 & 8.66 & 8.65 & 13.49 \\ 
			&   & 20 & 0.08 (0.08)& -0.23 & -0.55 & -0.55 & -0.00 & 7.07 (7.05) & 7.02 & 7.03 & 7.02 & 8.98 \\ 
			\midrule
			h2 & 1.00 & 10 & 0.44 & 1.44 & 0.95 & 224.52 & 25.76 & 21.13 & 21.54 & 21.51 & 226.07 & 70.23 \\ 
			&   & 20 & 0.24 & 1.20 & 0.85 & 222.61 & 11.63 & 19.73 & 20.03 & 20.01 & 224.24 & 44.09 \\ 
			& 2.50 & 10 & 0.17 & 0.25 & -0.28 & 77.71 & 10.02 & 13.02 & 13.14 & 13.14 & 79.11 & 33.13 \\ 
			&   & 20 & 0.02 & 0.19 & -0.16 & 77.31 & 4.29 & 11.78 & 11.89 & 11.89 & 78.59 & 21.82 \\ 
			& 5.00 & 10 & 0.08 & -0.21 & -0.57 & 42.28 & 5.61 & 10.10 & 10.04 & 10.05 & 43.58 & 21.16 \\ 
			&   & 20 & -0.01 & -0.15 & -0.36 & 42.08 & 2.41 & 8.58 & 8.54 & 8.54 & 43.10 & 14.14 \\ 
			\midrule
			h3 & 1.00 & 10 & 0.10 & -0.31 & -0.84 & -26.12 & -2.56 & 15.01 & 15.06 & 15.10 & 30.32 & 35.41 \\ 
			&   & 20 & 0.13 & -0.20 & -0.63 & -26.05 & -0.99 & 13.71 & 13.78 & 13.79 & 29.85 & 24.84 \\ 
			& 2.50 & 10 & 0.03 & -0.54 & -1.17 & -23.10 & -2.70 & 11.26 & 11.40 & 11.44 & 25.73 & 22.60 \\ 
			&   & 20 & -0.00 & -0.40 & -0.88 & -22.89 & -1.33 & 9.83 & 9.93 & 9.95 & 25.03 & 15.58 \\ 
			& 5.00 & 10 & 0.16 & -0.38 & -0.84 & -18.61 & -2.29 & 9.35 & 9.29 & 9.31 & 20.87 & 16.09 \\ 
			&   & 20 & 0.11 & -0.26 & -0.59 & -18.35 & -1.09 & 7.77 & 7.72 & 7.74 & 20.06 & 11.20 \\ 
			\midrule
			h4 & 1.00 & 10 & -0.07 & -0.61 & -0.66 & -82.81 & -9.07 & 5.62 & 5.82 & 5.84 & 83.06 & 19.03 \\ 
			&   & 20 & -0.07 & -0.56 & -0.63 & -70.24 & -4.13 & 5.30 & 5.47 & 5.49 & 70.51 & 12.21 \\ 
			& 2.50 & 10 & -0.07 & -0.66 & -0.79 & -82.18 & -9.13 & 6.40 & 6.49 & 6.52 & 82.47 & 21.13 \\ 
			&   & 20 & -0.05 & -0.58 & -0.75 & -70.05 & -3.99 & 6.05 & 6.13 & 6.16 & 70.34 & 13.60 \\ 
			& 5.00 & 10 & 0.12 & -0.26 & -0.42 & -80.77 & -8.95 & 6.73 & 6.76 & 6.78 & 81.08 & 21.85 \\ 
			&   & 20 & 0.12 & -0.21 & -0.42 & -69.06 & -3.99 & 6.32 & 6.34 & 6.36 & 69.38 & 14.16 \\ 
			\bottomrule
		\end{tabular}
	}
	\caption{RB($\%$) and RRMSE($\%$) for EB, EB\_clsd, EB\_HZ, $M$,PI,Dir of considred parametrs by sample size under the gamma mixed model} 
	\label{RB_RE_GMM}
\end{table}


\subsection{Simulation 2}

In Simulation 1, we observed that the empirical best predictors for the gamma-gamma model perform well even when the model is misspecified. Thus, we concentrate on the performance of the MSE estimators under the framework of the gamma-gamma model. We simulate data from the gamma-gamma model. The predictors and MSE estimators are cunstructed under the assumptions of the gamma-gamma model. 

We evaluate the MSE estimators described in Section 3.2 on the basis of two criteria. The first is a measure of the relative bias of the MSE estimator, as a measure of the unconditional MSE of the predictor. This is defined as 
\begin{align}\label{RB_MSE}
RB_{A}^{\rm uncond} = \frac{(10000D)^{-1}\sum_{i=1}^{D}\sum_{m=1}^{10000}{\rm mse}_{i}^{(m, A)}-{\rm MSE^{UCond}}}{{\rm MSE^{UCond}}},
\end{align}
where  ${\rm mse}_{i}^{(m, A)}$ is the type $A$ MSE estimator obtained in MC simulation $m$, \newline $A\in  \{\rm noBC, Add, Mult, HM, Comp, S, D\}$ and ${\rm MSE}^{\rm UCond} = {(10000D)}^{-1}\sum_{i=1}^{D}\sum_{m=1}^{10000}(\hat{\theta}_{i}^{\rm EB(m)}-{\theta}_{i}^{(m)})^{2}$. Note that the proposed estimators can also be regarded as the estimators for the conditional MSE defined as $E\{(\hat{\theta}_{i}^{EB}-{\theta}_{i})^{2}\mid\bm{y}_{is}\}$. Thus, we define the conditional RB  as 
\begin{align}\label{RB_MSE_Cond}
RB_{A}^{\rm cond} = \frac{(10000D)^{-1}\sum_{i=1}^{D}\sum_{m=1}^{10000}{\rm mse}_{i}^{(m, A)}-{\rm MSE^{\rm cond}}}{{\rm MSE^{\rm cond}}},
\end{align}
where 
\begin{align}
{\rm MSE}^{\rm Cond} &=D^{-1}\sum_{i=1}^{D}( \bar{M}_{1i}+\bar{M}_{2i}),  
\end{align}
 $\bar{M}_{1i} =(10000)^{-1}\sum_{m=1}^{M}(\hat{\theta}_{i}^{{\rm B}(m)}-\hat{\theta}_{i}^{(m)})^{2}$ and $\bar{M}_{2i} =(10000)^{-1}\sum_{m=1}^{M}(\hat{\theta}_{i}^{{\rm EB}(m)}-\hat{\theta}_{i}^{{\rm B}(m)})^{2}$. \cite{lohr2009jackknife} evaluate the conditional relative bias of the MSE estimators in their simulations, and \cite{booth1998standard} discuss the value of the conditional MSE estimators in prediction problems. 

Several of the MSE estimators defined in Section 3 incorporate corrections for the estimators of the bias of the estimator of the leading term. We therefore check whether there exists a bias for the leading term estimators. For this, define a test-statistic as
\begin{align}\label{leadtest}
T^{\rm Bias} = \frac{\bar{\omega}}{{\rm sd}_{\omega}/\sqrt{10000}},
\end{align}
where $\omega^{(m)}=D^{-1}\sum_{d=1}^{D}(\hat{M}_{1i}^{(m)}-{M}_{1i}^{(m)})$, $\bar{\omega}=10000^{-1}\sum_{m=1}^{10000}\omega^{(m)}$ and\\ ${\rm sd}_{\omega} =\sqrt{(10000-1)^{-1}\sum_{m=1}^{10000}(\omega^{(m)}-\bar{\omega})^2}$. 

\begin{figure}[H]
	\centering
	\includegraphics[width=15cm]{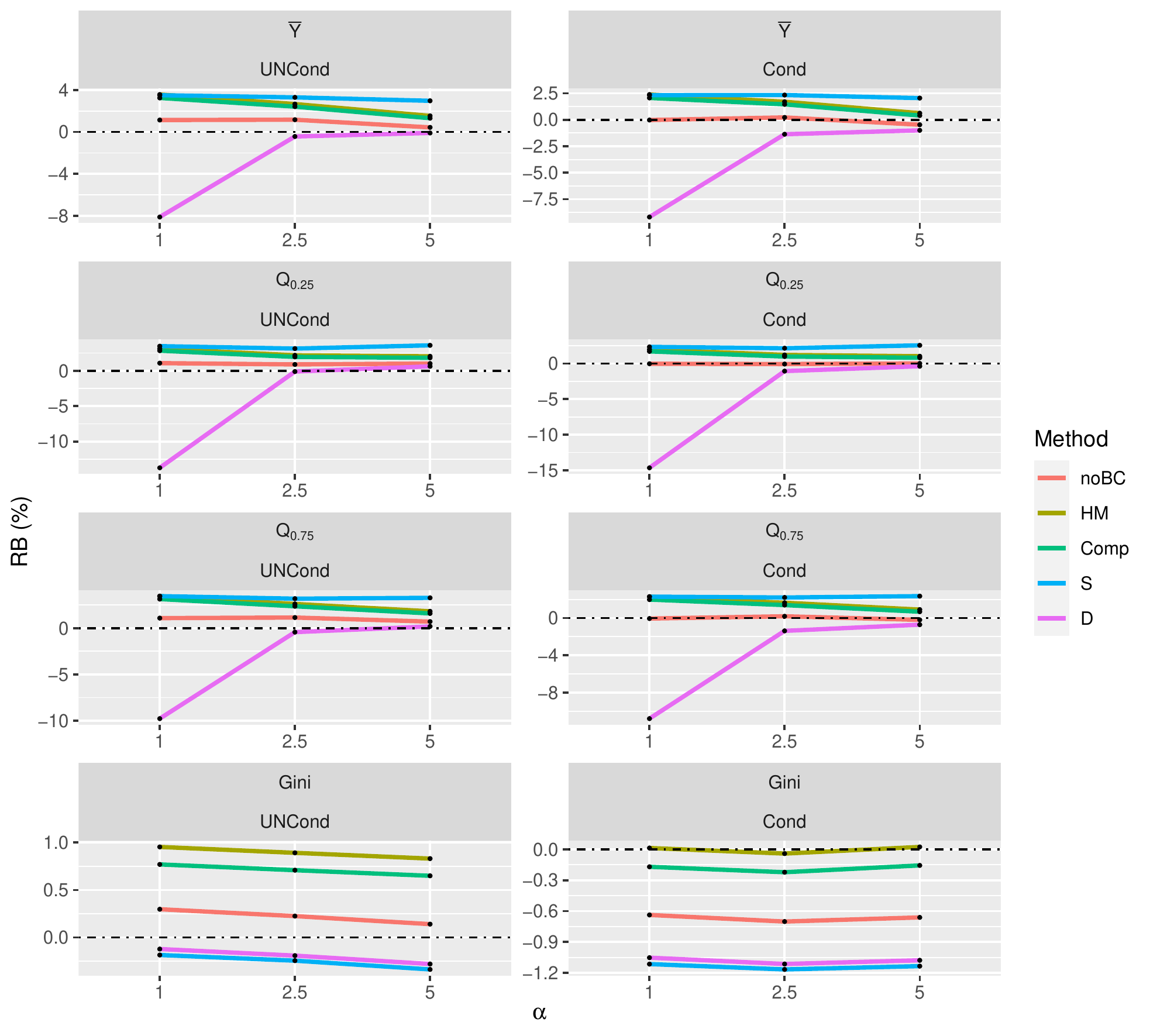}
	\caption{RB (\%) of MSE Estimators}
	\label{Sim2}
\end{figure}

Figure~\ref{Sim2} displays the relative biases of the alternative MSE estimators. The single-bootstrap MSE estimator ($S$) has a positive bias for $\bar{Y}$, $Q_{0.25}$, and $Q_{0.75}$. The double-bootstrap procedure ($D$) can over-correct this bias, producing important negative biases when $\alpha = 1$. For Gini, both $S$ and $D$ have negative biases. The MSE estimator noBC consistently has relative bias close to zero. The bias corrections Comp and HM lead to slight increases in the estimated MSE. The t-statistics in Table~\ref{BiasTeststat} shed insight into the relative biases of the noBC, Comp, and HM MSE estimators. The estimator of the leading term does not have a significant bias for all parameters, except for the Gini coefficient. Therefore, the noBC MSE estimator has RB close to zero for  $\bar{Y}$, $Q_{0.25}$, and $Q_{0.75}$. For the Gini coefficient, the bias of the estimator of the leading term is important. As illustrated by the conditional relative bias of the MSE estimators, the HM and Comp bias corrections effectively correct the bias of the estimator of the leading term for the Gini coefficient. 

\begin{table}[H]
	\centering
	\begin{tabular}{c|ccc}
		\toprule\toprule
		\multirow{2}{*}{Parameter}&\multicolumn{3}{c}{$\alpha$}\\
		\cline{2-4}
		& 1 & 2.5 & 5 \\ \hline
		h1 & -0.13 & 0.35 & -0.96 \\ 
		h2 & -0.08 & -0.15 & -0.03 \\ 
		h3 & -0.20 & 0.29 & -0.45 \\ 
		h4 & -4.62 & -5.60 & -4.85 \\ 
		\bottomrule\bottomrule
	\end{tabular}
	\caption{ $T^{\rm Bias}$ by the area parameter for each scenario} 
	\label{BiasTeststat}
\end{table}

\subsection{Simulation 3}

In this simulation study, the setup described in Section 5.1.1 is slightly modified to take into account an informative sample design. Instead of using simple random sampling when selecting $n_{i}$ units within an area $i$, we use systematic sampling. The inclusion probability for element $j$ within area $i$ is defined as 
$$\pi_{ij} =
\frac{n_{i}\exp(ax_{ij}+by_{ij}+\tau_{ij}/20)}{\sum_{k=1}^{N_{i}}\exp(ax_{ik}+by_{ik}+\tau_{ik}/20)},$$
where $\tau_{ij}\stackrel{iid}{\sim}\text{Gamma}(\delta,\delta)$. We hold $a$ constant at $0.05$. We vary $b$ from $0.05$ to $0.2$ with an increment of 0.05 to alter the degree of the informativeness in the sample design. Then, the performance of the alternative predictors is examined through the RB and RRMSE defined in Section 5.1.1. 

Table~\ref{sim3_RB_RE} contains the average RBs and RRMSEs of the alternative predictors, where the average is across areas of the same sample size.  EB\_INFO always attains smallest RBs and RRMSEs for all scenarios. The RBs of the EB, EB\_HZ, and $M$, that ignore the unequal sample probabilities, significantly increase as $b$ increases. Among these predictors, $M$ performs best in terms of both RB and RRMSE. EB was robust to the model misspecification, but seems to be more sensitive to the informative sampling strategy, as shown by its bigger RBs. Fortunately, the EB-Info procedure offers a theoretically defensible solution for the informative sample design.  



\begin{sidewaystable}[ht]
	\centering
	\scalebox{0.8}{
		\begin{tabular}{ccc|cccccc|cccccc}
			\toprule \toprule
			\multirow{2}{*}{\large{\bfseries Parameter }} &\multirow{2}{*} {\large{\bfseries $\alpha$}} & \multirow{2}{*}{\large{\bfseries $b$}} & \multicolumn{6}{c}{\large{\bfseries{RB (\%)}}}& \multicolumn{6}{c}{\large{\bfseries{RRMSE (\%)}}}\\\cmidrule{4-15}
			& & &{\large{\bfseries EB\_INFO}} & {\large{\bfseries EB}} & {\large{\bfseries EB\_HZ}} & {\large{\bfseries M}} & {\large{\bfseries PI}} & {\large{\bfseries Dir}} & {\large{\bfseries EB\_INFO}} & {\large{\bfseries EB}} & {\large{\bfseries EB\_HZ}} & {\large{\bfseries M}} & {\large{\bfseries PI}} & {\large{\bfseries Dir}} \\ 
			\midrule
			h1 & 1.00 & 0.05 & -0.01 & 2.58 & 0.60 & -2.77 & -2.77 & 1.97 & 27.29 & 28.50 & 31.41 & 27.54 & 27.53 & 35.84 \\ 
			&  & 0.10 & -0.13 & 5.21 & 2.98 & -0.36 & -0.36 & 4.90 & 26.41 & 30.75 & 30.59 & 26.90 & 26.89 & 40.89 \\ 
			&  & 0.20 & -0.42 & 10.84 & 8.12 & 4.73 & 4.73 & 11.34 & 25.85 & 36.58 & 31.51 & 28.52 & 28.50 & 55.51 \\ 
			& 5.00 & 0.05 & -0.03 & 2.55 & -2.82 & -0.75 & -0.75 & 2.81 & 13.03 & 15.13 & 31.11 & 13.82 & 13.82 & 21.30 \\ 
			&  & 0.10 & -0.13 & 5.17 & -0.75 & 1.66 & 1.66 & 6.67 & 13.05 & 18.61 & 30.33 & 13.42 & 13.42 & 28.43 \\ 
			&  & 0.20 & -0.64 & 9.97 & 3.54 & 6.07 & 6.07 & 14.08 & 13.51 & 23.67 & 27.19 & 15.29 & 15.28 & 38.65 \\ 
			\midrule
			h2 & 1.00 & 0.05 & 0.13 & 2.38 & 4.36 & 1.49 & 147.83 & 21.20 & 34.04 & 35.03 & 38.59 & 34.30 & 178.81 & 76.84 \\ 
			&  & 0.10 & 0.12 & 4.78 & 6.66 & 3.86 & 152.29 & 24.42 & 33.50 & 37.06 & 38.77 & 34.63 & 188.27 & 88.72 \\ 
			&  & 0.20 & 0.04 & 9.92 & 11.60 & 8.83 & 161.54 & 32.29 & 33.43 & 42.68 & 40.29 & 37.83 & 206.04 & 141.07 \\ 
			& 5.00 & 0.05 & 0.12 & 2.19 & -5.32 & -2.70 & 18.32 & 7.69 & 14.86 & 16.58 & 36.88 & 16.20 & 25.50 & 31.35 \\ 
			&  & 0.10 & 0.14 & 4.44 & -3.76 & -0.73 & 20.44 & 11.51 & 14.96 & 19.69 & 36.39 & 15.47 & 28.48 & 43.40 \\ 
			&  & 0.20 & -0.03 & 8.64 & -0.08 & 3.03 & 24.40 & 19.73 & 15.22 & 24.23 & 31.67 & 15.99 & 33.40 & 62.77 \\ 
			\midrule
			h3 & 1.00 & 0.05 & -0.00 & 2.48 & 0.47 & -2.67 & -11.30 & 0.38 & 28.80 & 29.98 & 33.56 & 29.06 & 32.41 & 41.65 \\ 
			&  & 0.10 & -0.13 & 4.99 & 2.71 & -0.38 & -8.93 & 3.23 & 28.17 & 32.38 & 33.03 & 28.70 & 30.32 & 47.63 \\ 
			&  & 0.20 & -0.34 & 10.49 & 7.69 & 4.60 & -3.80 & 9.72 & 27.77 & 38.32 & 33.60 & 30.51 & 28.47 & 62.14 \\ 
			& 5.00 & 0.05 & -0.04 & 2.61 & -2.59 & -0.36 & -3.90 & 1.18 & 14.03 & 16.09 & 33.60 & 14.72 & 16.33 & 25.02 \\ 
			&  & 0.10 & -0.16 & 5.29 & -0.50 & 2.13 & -1.41 & 5.15 & 14.15 & 19.61 & 33.03 & 14.57 & 14.88 & 31.19 \\ 
			&  & 0.20 & -0.69 & 10.28 & 3.97 & 6.74 & 3.19 & 12.93 & 14.67 & 24.83 & 29.21 & 16.78 & 15.06 & 40.09 \\ 
			\midrule
			h4 & 1.00 & 0.05 & -0.00 & 0.11 & -1.33 & -1.64 & -55.40 & -6.83 & 5.58 & 5.59 & 5.81 & 5.89 & 55.74 & 15.78 \\ 
			&  & 0.10 & -0.01 & 0.23 & -1.24 & -1.56 & -54.96 & -6.76 & 5.57 & 5.58 & 5.77 & 5.85 & 55.31 & 15.71 \\ 
			&  & 0.20 & -0.03 & 0.46 & -1.05 & -1.37 & -54.05 & -6.59 & 5.54 & 5.57 & 5.70 & 5.77 & 54.40 & 15.60 \\ 
			& 5.00 & 0.05 & -0.01 & 0.45 & 3.39 & 2.82 & -35.65 & -6.42 & 6.27 & 6.29 & 7.65 & 6.89 & 36.23 & 17.62 \\ 
			&  & 0.10 & -0.05 & 0.89 & 3.91 & 3.29 & -34.81 & -6.09 & 6.24 & 6.30 & 7.95 & 7.06 & 35.40 & 17.50 \\ 
			&  & 0.20 & -0.25 & 1.61 & 4.56 & 3.94 & -33.34 & -5.68 & 6.26 & 6.43 & 8.09 & 7.36 & 33.94 & 17.42 \\ 
			\bottomrule
		\end{tabular}
	}
	\caption{RB ($\%$) and RRMSE ($\%$) for EB, EB\_{HZ}, M, PI, Dir of considered parameters by sample size under the informative sample design} 
	\label{sim3_RB_RE}
\end{sidewaystable}

\section{Application to Ohio Soil Erosion Data}

This analysis is based on a survey of cropland conducted as part of the Conservation Effects Assessment Project (CEAP), which is a nationwide evaluation to quantify the effectiveness of conservation efforts on croplands. The sample for the CEAP survey contains a subset of crop fields in a massive panel survey called the National Resources Inventory (NRI). The NRI measures numerous variables related to  land cover/use and soil characteristics on non-federal US lands. One variable of interest is sheet and rill erosion, soil loss due to rainfall or water runoff. 


There are two measurements of the soil erosion: the Universal Soil Loss Equation (USLE) and Revised Universal Soil Loss Equation version 2 (RUSLE2). The former is a classic soil erosion measurement (See Wischmeier and Smith, 1965) and the latter is the enhaced version of the former in that it considers additional variables and daily-based factors. USLE is obtained at every NRI sample point. The measure of  RUSLE2 that we use is only computed for  the CEAP subsample. Our objective is to estimate functions of RUSLE2 for Ohio counties using the USLE as a covariate.

Let $y_{ij}$ and $x_{ij}$ denote the RUSLE2 and USLE for the $j$th sample point of the $i$th county in Ohio. Note that 73 counties among 88 counties are sampled, and nonsampled counties are excluded in this analysis. Further, we regress $log(1/\pi_{ij})$ on $x_{ij}$ and $y_{ij}$ with areas fixed effects. The coefficients for both $a$ and $b$  were not significant, so we do not consider the predictor under an informative sample design. We fit the gamma-gamma model (\ref{Gam-Gam}) and the GLMM (\ref{Hobza}) to the sample. 

\begin{figure}[H]
	\centering
	\includegraphics[width = 16cm]{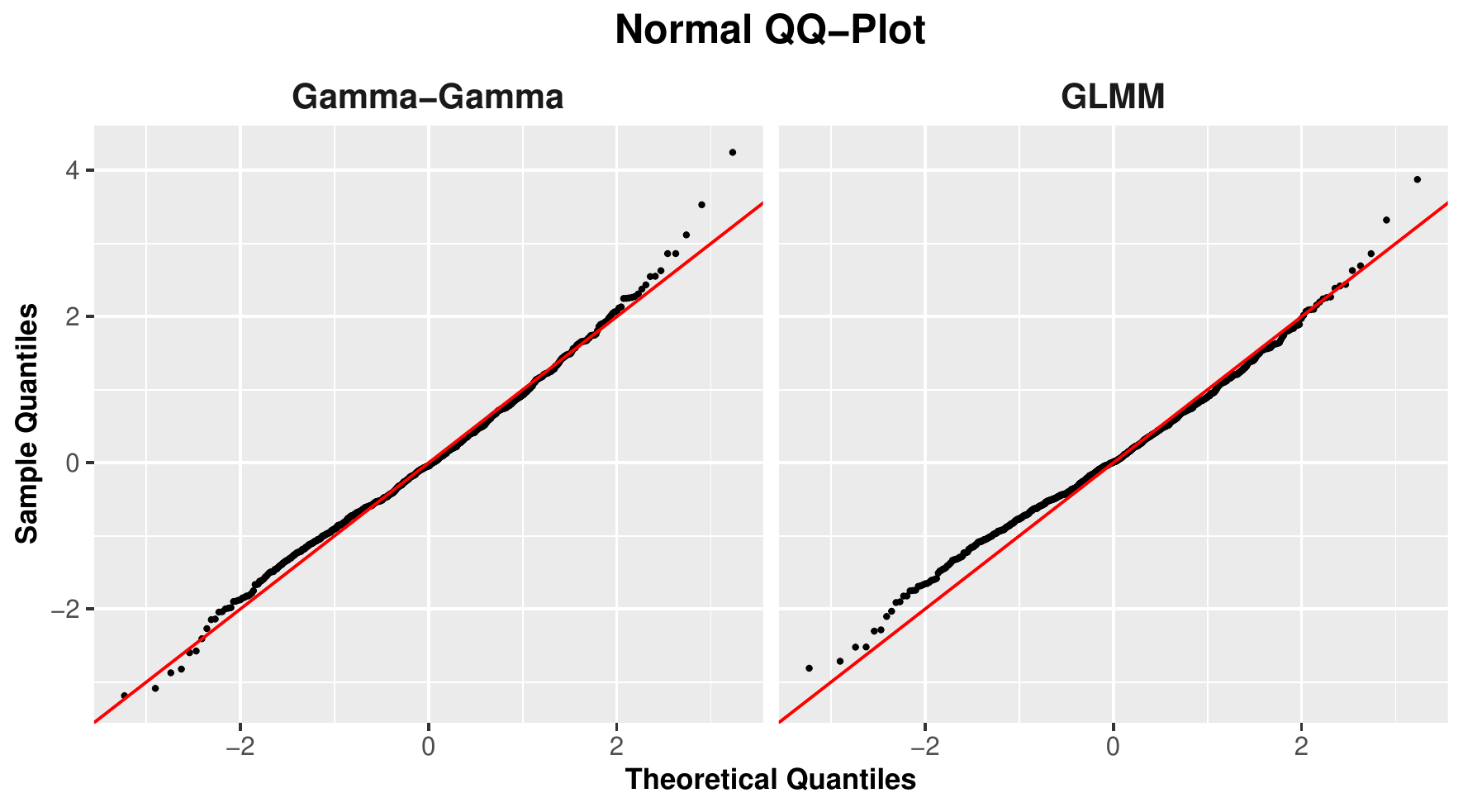}
	\caption{Generalized residuals of the gamma-gamma model (Left) and the GLMM (Right)}
	\label{gen_res}
\end{figure}

The fitted models are assessed by using the generalized residuals. The gneralized residual for the gamma-gamma model is defined as 
$r_{ij} = G^{-1}(y_{ij}\mid \bm{x}_{ij}, \hat{u}_{i};\hat{\bm{\psi}})$, where $G$ is the CDF of a gamma distribution with shape parameter $\hat{\alpha}$ 
 and rate parameter $\mbox{exp}(\bm{x}_{ij}'\hat{\bm{\gamma}})\hat{u}_{i}$ with 
 $$\hat{u}_{i}= \frac{ \hat{\alpha} + \hat{\delta}}{ \sum_{j = 1}^{n_{i}}y_{ij}\mbox{exp}(\bm{x}_{ij}'\hat{\bm{\gamma}}) + \hat{\tau}  }.$$  For the GLMM, the generalized residual is defined as $r_{ij} = G^{-1}(y_{ij}\mid \bm{x}_{ij}, \hat{u}_{i};\hat{\bm{\psi}}^{GLMM})$, where $G$ is the CDF of a gamma distribution with shape parameter $\hat{\nu}$ and rate parameter $\hat{\nu}^{-1}\mbox{exp}(-\bm{x}_{ij}'\hat{\bm{\beta}} + \hat{v}_{i})$ with $\hat{v}_{i}$ being the predicted random effect from {\tt ranef} in {\tt R}.  The set of the generalized residuals $\{r_{ij}:j=1,\hdots,n_{i}, i=1,\hdots,73\}$ behaves like a sample from the uniform distribution under the correct model. The justification for the fitted model can be made through comparing the distribution of $\{\Phi^{-1}(r_{ij}):j=1,\hdots,n_{i}, i=1,\hdots,73\}$ with the normal distribution. As shown in Figure \ref{gen_res}, the residuals from both models almost fall along the 45 degree reference line. Both the GLMM and gamma-gamma models appear to fit the data adequately. 

\begin{figure}[H]
	\centering
	\includegraphics[width=15cm]{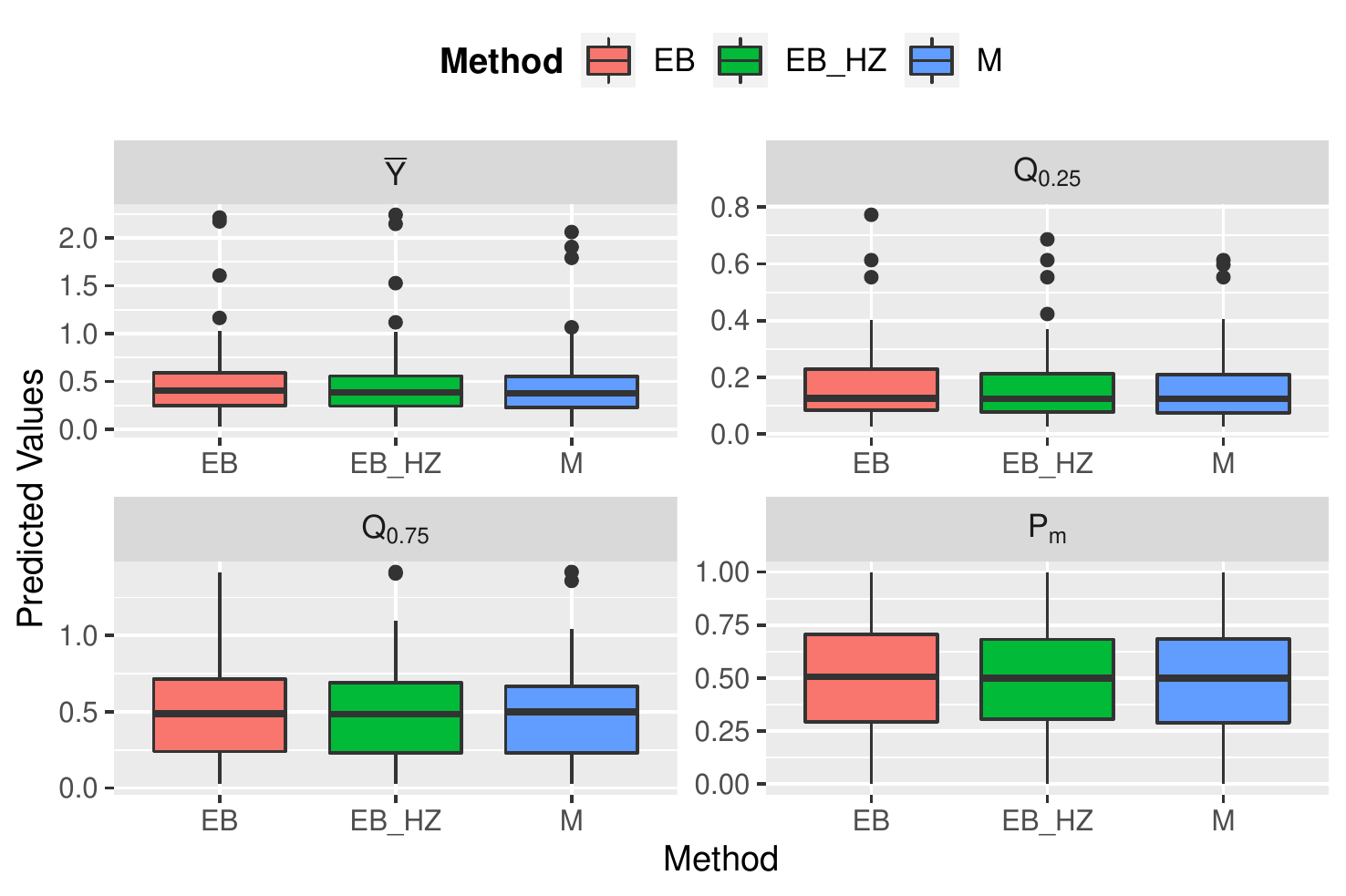}
	\caption{Predicted values of the considered parameters by EB, EB\_HZ, M}
	\label{data_pred}
\end{figure}

We next consider the following county-level parameters: the mean ($\bar{Y}_{i}$), the $25$th ($Q_{0.25})$ \& the $75$th ($Q_{0.75}$) quantiles, and the proportion greater than $0.22$ ($P_{m}$) of a county $i$, $i=1,\hdots,73$. The value used for $P_{m}$ is the sample median of RUSLE2 in Ohio. Based on the simulation results in Section 5.1.1, we take into account EB, EB\_HZ, and M in Section 3. As shown in Figure \ref{data_pred}, the predicted values produced by each predictor are distributed similarly for a given parameter.  We construct normal theory 95\% confidence intervals for the EB county predictors using the noBC, HM, and S MSE estimators. The double-bootstrap produced negative MSE estimators for some counties, so this procedure cannot be used to construct CIs.  CIs constructed with S tend to be longer than CIs constructed by noBC and HM. This reflects the result of the simulation study where the $S$ MSE estimator tended to have a positive bias. 


\begin{table}[H]
	\centering
	\begin{tabular}{cc|cc}
		\toprule
		$\hat{\bm{\psi}}^{\rm Gam-Gam}$ & Estimates  & $\hat{\bm{\psi}}^{\rm GLMM}$ & Estimates \\ 
		\hline
		$\hat{\alpha}$ & 1.659 (1.617, 1.700) & $\hat{\nu}$& 1.381 (1.224, 1.539) \\ 
		$\hat{\delta}$ & 4.922 (3.931, 5.912) & $\hat{\phi}^{2}$ &0.458 (0.370, 0.546) \\ 
		$\hat{\gamma}_{0}$ & 2.183 (0.365, 4.003) & $\hat{\beta}_{0}$&-1.602 (-1.734, -1.471)\\ 
		$\hat{\gamma}_{1}$ & -0.156 (-0.19, -0.126) & $\hat{\beta}_{1}$& 0.153 (0.139, 0.168)  \\ 
		\bottomrule
	\end{tabular}
	\caption{Estimates of the model parameters with their confidence intervals}
	\label{model_par_est}
\end{table}

\begin{figure}[H]
	\centering%
	\includegraphics[width = 18cm]{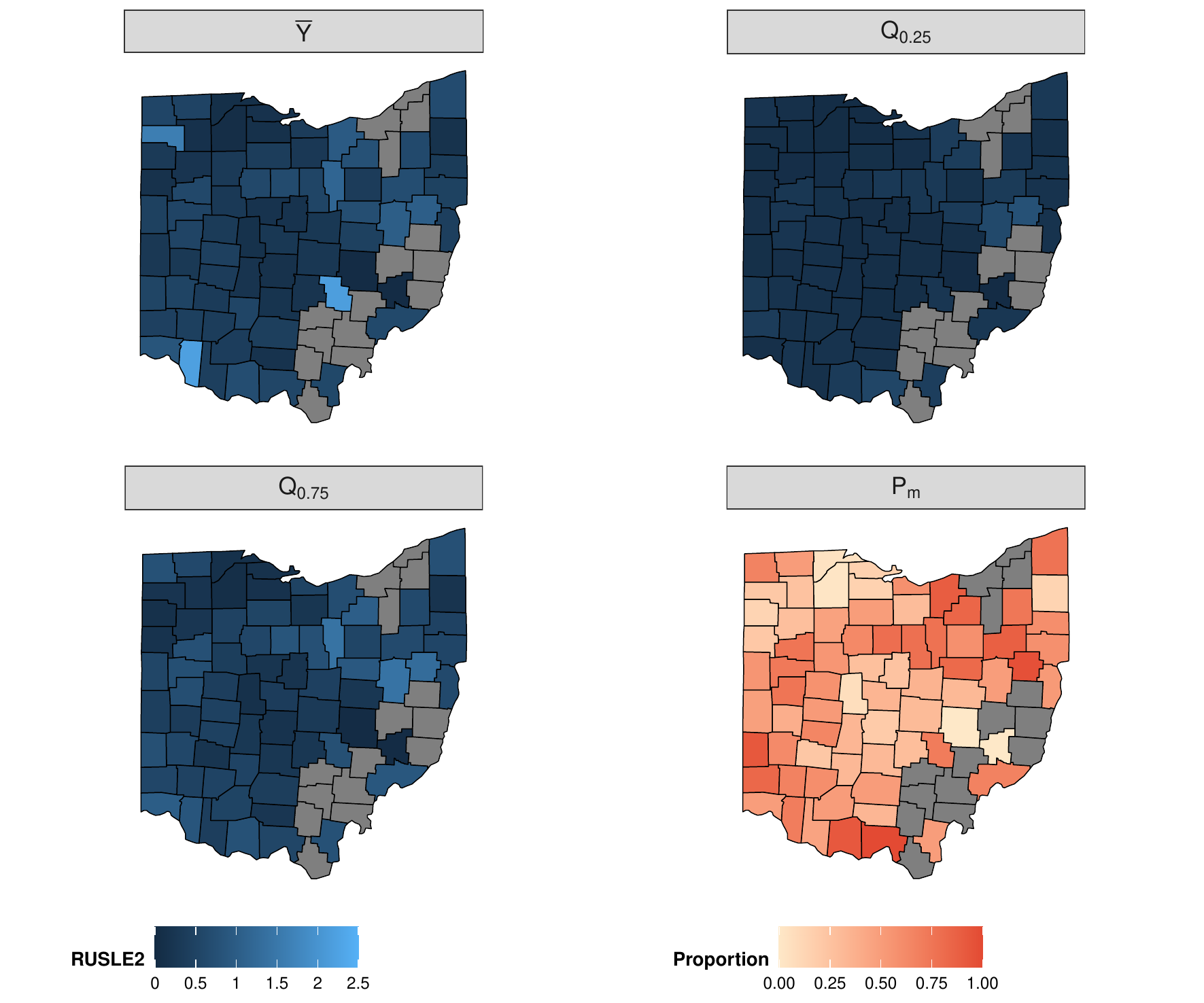}
	\caption{Prediction of the considered small area parameters with EB predictor. Counties colored by grey are nonsampled.}
	\label{county_plot}
\end{figure}


\begin{figure}[H]
	\centering
	\includegraphics[width = 16cm]{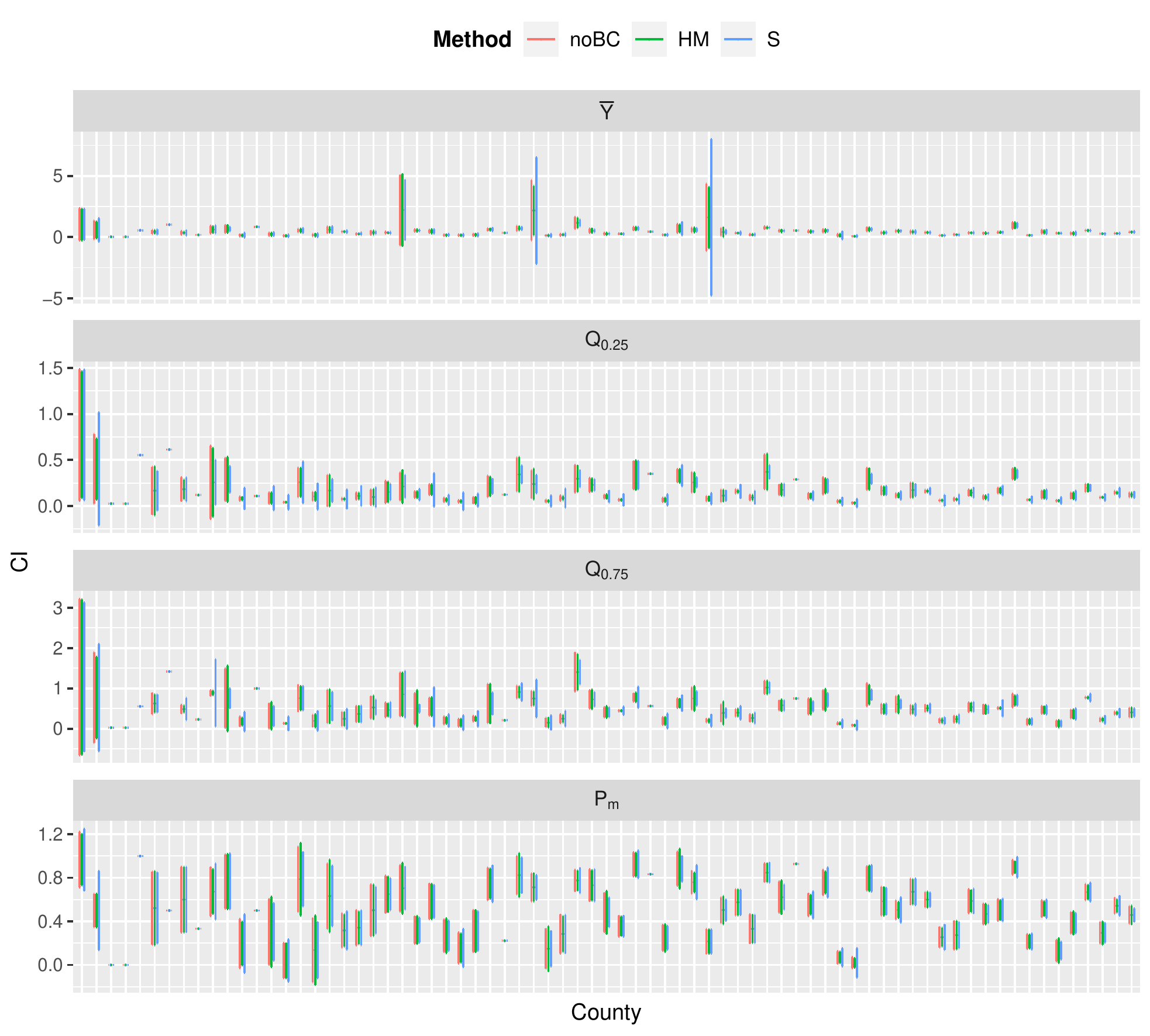}
	\caption{95 \% normal-theory Confidence Intervals. Counties are sorted by their sample size.}
	\label{data_CI}
\end{figure}

\section{Conclusion}

\hspace{.2 in }  In this work, we demonstrate that the gamma distribution is a useful model for constructing small area predictors with a skewed response variable. We focus heavily on the EB predictor for the gamma-gamma model proposed in \cite{Graf2019}. We demonstrate that this predictor has good properties, even when the true model is the gamma-GLMM. We also demonstrate that the general MSE estimator of \cite{https://doi.org/10.48550/arxiv.2210.12221} has good properties for in the context of the gamma-gamma model. Finally, we extend the gamma-gamma model to an informative sample design. Our results generally provide more support for the gamma-gamma model than for the gamma-GLMM.


\bibliographystyle{agsm}

\bibliography{References}

\appendix

\section{Proof of Theorem 1}

Without loss of generality, for $i=1, \hdots, D$, let the first $n_{i}$ units in $\bm{y}_{i}=(y_{i1},\hdots,y_{iN_{i}})^{T}$ be sampled and the others non-sampled, denoted as $\bm{y}_{is} = (y_{i1},\hdots,y_{in_{i}})^{T}$ and then $\bm{y}_{ir} = (y_{in_i+1},\hdots,y_{iN_{i}})^{T}$, respectively. Then, the best predictor of the $i$th area mean boils down to deriving the conditional expectation of an area random effect given observed units in the $i$th area, $\bm{y}_{is}$, by the following equations: 
 \begin{align*}
       {\bar{y}}_{N_i}^{BP} 
       &= E(\bar{y}_{N_{i}}\mid\bm{y}_{is})\\
         &= \frac{1}{N_i}\bigg\{\sum_{j=1}^{n_i}y_{ij}+\sum_{j=n_{i}+1}^{N_i}E\big[y_{ij}\mid\bm{y}_{is}\big]\bigg\}\\
      &= \frac{1}{N_i}\bigg\{\sum_{j=1}^{n_i}y_{ij}+\sum_{j=n_{i}+1}^{N_i}E\big[E[y_{ij}\mid u_{i}]\mid\bm{y}_{is}\big]\bigg\} \text{\hspace{0.1in}$\because y_{ik}\perp y_{il} \mid u_{i}$}\\
              &= \frac{1}{N_i}\bigg\{\sum_{j=1}^{n_i}y_{ij}+\sum_{j=n_{i}+1}^{N_i}\alpha \exp(-\bm{x}_{ij}^{T}\bm{\gamma})E[u_{i}^{-1}\mid\bm{y}_{is}]\bigg\}. 
 \end{align*}
 
\hspace{.2in}When deriving the conditional distribution of the area random effect, we can again take advantage of the parameterization and the conjugate pair in model (\ref{Gam-Gam}), obtaining the kernel of a gamma distribution with shape parameter $n_{i}\alpha+\delta$ and rate parameter $\big\{\sum_{j=1}^{n_{i}}y_{ij}\exp(\bm{x}_{ij}^{T}\bm{\gamma})\big\}+\delta$:  
\begin{align}\label{cond_ui}
      f(u_{i}\mid \bm{y}_{is}) &\propto f(\bm{y}_{is}\mid u_{i})\times f(u_{i})\\
      &=\bigg\{\prod_{j=1}^{n_i}f({y}_{ij}\mid u_{i})\bigg\}\times f(u_{i})\\
      &\propto u_{i}^{n_{i}\alpha+\delta-1}\exp\Bigg(-\bigg(\sum_{j=1}^{n_{i}}\{y_{ij}\exp(\bm{x}_{ij}^{T}\bm{\gamma})\}+\delta\bigg)u_{i}\Bigg).
  \end{align}
Therefore, using the inverse-gamma distribution property, we can calculate
\begin{align}
   E[u_{i}^{-1}\mid\bm{y}_{is}]&=\frac{\big\{\sum_{j=1}^{n_{i}}y_{ij}\exp(\bm{x}_{ij}^{T}\bm{\gamma})\big\}+\delta}{n_{i}\alpha+\delta-1}, 
\end{align}

\section{Proof of Theorem 2}

\begin{align*} 
        f(\bm{y}_{is}) &=\int f(\bm{y}_{is}\mid u_{i})f(u_{i})du_{i}\\
        &=\int\frac{\exp\big(\alpha (\sum_{j=1}^{n_i}\bm{x}_{ij})^{T}\bm{\gamma}\big)u_{i}^{n_{i}\alpha}}{\{\Gamma(\alpha)\}^{n_{i}}}\prod_{j=1}^{n_i}y_{ij}^{\alpha-1}\exp\Big(-\sum_{j=1}^{n_{i}}\{y_{ij}\exp(\bm{x}_{ij}^{T}\bm{\gamma})\}u_{i}\Big)\times\frac{\delta^{\delta}}{\Gamma(\delta)}u_{i}^{\delta-1}\exp(-u_{i}\delta)du_{i}\\
        &=\frac{\exp\big(\alpha (\sum_{j=1}^{n_i}\bm{x}_{ij})^{T}\bm{\gamma}\big)}{\{\Gamma(\alpha)\}^{n_{i}}}\prod_{j=1}^{n_i}y_{ij}^{\alpha-1}\frac{\delta^{\delta}}{\Gamma(\delta)}\int u_{i}^{n_{i}\alpha+\delta-1}\exp\Bigg(-\bigg(\sum_{j=1}^{n_{i}}\{y_{ij}\exp(\bm{x}_{ij}^{T}\bm{\gamma})\}+\delta\bigg)u_{i}\Bigg) du_{i}\\
        &=\frac{\delta^{\delta}}{\{\Gamma(\alpha)\}^{n_{i}}\Gamma(\delta)}\prod_{j=1}^{n_i}y_{ij}^{\alpha-1}\exp\bigg(\alpha (\sum_{j=1}^{n_i}\bm{x}_{ij})^{T}\bm{\gamma}\bigg)\frac{\Gamma(n_{i}\alpha+\delta)}{\bigg(\sum_{j=1}^{n_{i}}\{y_{ij}\exp(\bm{x}_{ij}^{T}\bm{\gamma})\}+\delta\bigg)^{n_{i}\alpha+\delta}}.
    \end{align*}

\end{document}